\documentclass[aps,prb,preprint,showpacs]{revtex4-1}
\usepackage{amsmath}
\usepackage{graphicx}
\usepackage{dcolumn}
\usepackage{amsmath}
\usepackage{amssymb}
\usepackage{graphicx}
\usepackage{color}

\begin{document}

\title{Quantum magnetotransport in 
a bilayer MoS$_{2}$: \\influence of a perpendicular electric  field}
\author{M. Zubair$^{1}$, M. Tahir$^{2,*}$, P. Vasilopoulos$^{3}$, and K. Sabeeh$^{1}$}
\affiliation{$^{1}$Department of Physics,Quaid-i-Azam University, Islamabad $45320$, Pakistan}
\affiliation{$^{2}$Department of Physics, College of Science, University of Hafr Al Batin,
P.O. Box 1803, Hafr Al Batin 31991, Kingdom of Saudi Arabia}
\affiliation{$^{3}$Department of Physics, Concordia University, Montreal, Quebec, Canada H3G 1M8}

\begin{abstract}
We first derive the energy dispersion of bilayer MoS$_{2}$ in the presence of a perpendicular electric field $E_z$. 
We show that the 
band gap and layer splitting can be controlled by the field $E_z$. Away from the $k$ point, the intrinsic SOC splitting increases in the conduction band but is weakly affected in the valence band. We then analyze the band structure in the presence of a perpendicular magnetic field $B$ and the field $E_z$, including spin and valley Zeeman terms, and 
evaluate the Hall and longitudinal conductivities. We discuss the numerical results as  functions of the  fields $B$ and  $E_z$ for finite  temperatures. 
The field $B$ gives rise to a significant spin splitting in the conduction band, to a beating in the Shubnikov-de Haas (SdH) oscillations when it's weak, and to their splitting when it's strong. The Zeeman terms and $E_{z}$ 
suppress the beating and change the positions of the beating nodes of the SdH oscillations at low $B$ fields and enhance their splitting at high $B$ fields. Similar beating patterns are observed in the spin and valley polarizations at low $B$ fields. Interestingly, a $90\%$ spin polarization and a $100\%$ 
  square-wave-shaped valley polarization are observed at high $B$ fields. The Hall-plateau sequence depends on $E_z$. 
These findings may be pertinent to future  spintronic and valleytronic devices.
\end{abstract}

\maketitle

\section{Introduction}
Recently the MoS$_{2}$ monolayer has provided a new testbed for the study of fermion physics in reduced dimensions. 
Its strong intrinsic SOC and huge band gap \cite{5i}, approximately  $2\lambda=150$ meV  and $2\Delta=1.66$ eV, respectively,  render  it  pertinent to potential applications in spintronics 
and optoelectronics \cite{1i, 2i,  3i, 4i}. 
 Due to these features, MoS$_{2}$ may be more appropriate for device applications than graphene and the conventional two-dimensional electron gas (2DEG). 
Other investigated properties of  monlayer MoS$_{2}$ are magnetocapacitance \cite{7ii}, spin- and valley-dependent magnetooptical spectra \cite{8i, 8ii, 8iii} 
and an unconventional quantum Hall effect (QHE) \cite{4}. Most recently,  
magnetotransport studies of monolayer MoS$_{2}$ have been carried out \cite{5, 9i, z8}.   

In addition to monolayer  MoS$_{2}$, 
it has been recently realized that bilayer MoS$_{2}$ has potential applications in optoelectronics and spintronics. Also, a band-gap tuning is possible in a MoS$_{2}$ bilayer in the presence of a perpendicular electric field $E_z$ \cite{6,7,8}. 
Additional 
reported properties of bilayer MoS$_{2}$ 
include magnetoelectric effects and valley-controlled spin-quantum gates \cite{9}, tuning of the valley magnetic moment \cite{10}, and electrical control of the valley-Hall effect \cite{11}. Moreover, a field-effect transistor has been realized experimentally  in a few-layer MoS$_{2}$ \cite{13}. In contrast, bilayer graphene has intrinsically a very weak SOC \cite{14, 15} and, when not biased, a zero band gap \cite{16, 17, 18}. There exist  numerous theoretical and experimental \cite{17, 19, 20, 21, 22} studies of magnetotransport properties in bilayer graphene.  Although its band gap  can be controlled by an 
electric field $E_z$ \cite{1b, 2b, 3b, 4b}, high-quality samples of MoS$_{2}$ bilayers with a strong intrinsic SOC and a huge band gap are of particular importance. Contrary to bilayer graphene,  the MoS$_{2}$ bilayer has greater potential for future spintronic and valleytronic applications. Recently,   not only the 
 QHE but also the SdH  oscillations have been observed in high-quality monolayer and multilayer MoS$_{2}$ \cite{23} 
but neither magnetotransport nor the effect of an electric field $E_z$ have, to our knowledge, been theoretically studied  for bilayer MoS$_{2}$.  
 Such a study is the aim of the present work. 
 
The paper is organized  as follows. In Sec. II we formulate the problem  and discuss the  band structure of  bilayer MoS$_{2}$ with the help of the eigenvalues, eigenfunctions, Fermi energy, and density of states (DOS). We then evaluate  the Hall and longitudinal conductivities using  the linear-response 
formulas of Ref. \onlinecite{24}.   Interestingly, we find that the Hall-plateau sequence depends on the field $E_z$ and  becomes unconventional 
 when  $E_z$   is present. 
 Also, we compare the results with those on bilayer graphene.  
Concluding remarks follow in Sec. IV.
\begin{figure}[t]
\centering

\includegraphics[width=.32\textwidth]{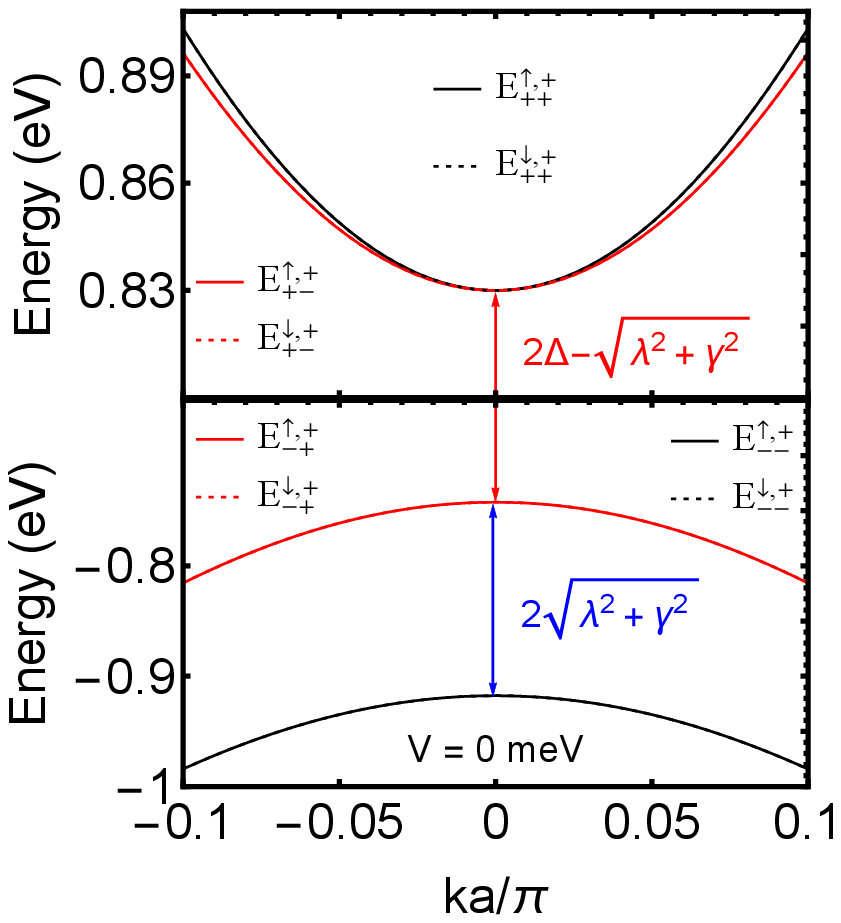}
\hspace*{0.9cm}
\includegraphics[width=.273\textwidth]{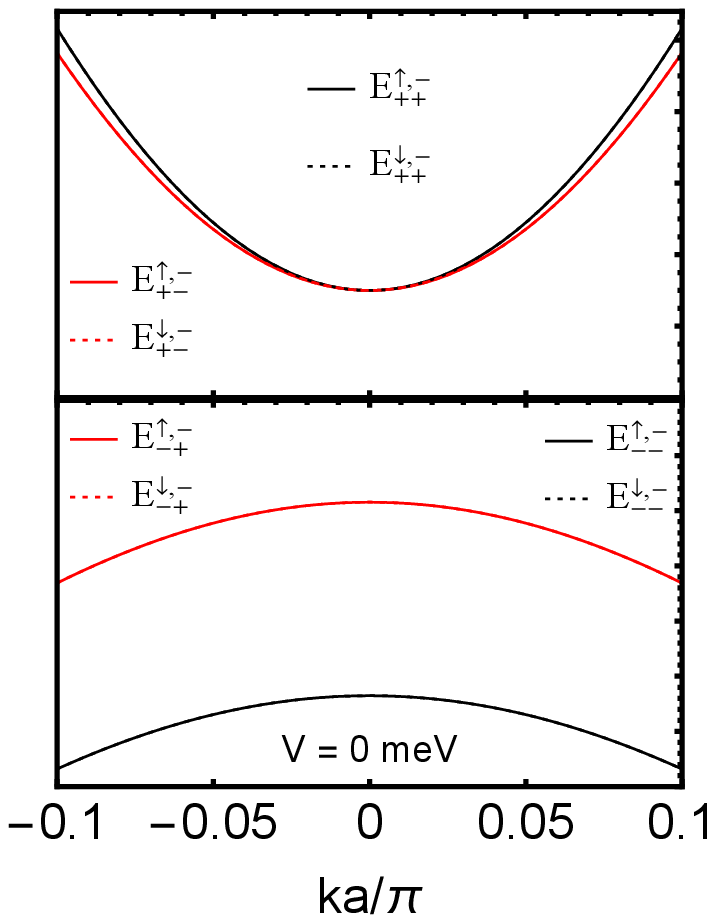}
\ \\
\ \\
\centering
\includegraphics[width=.32\textwidth]{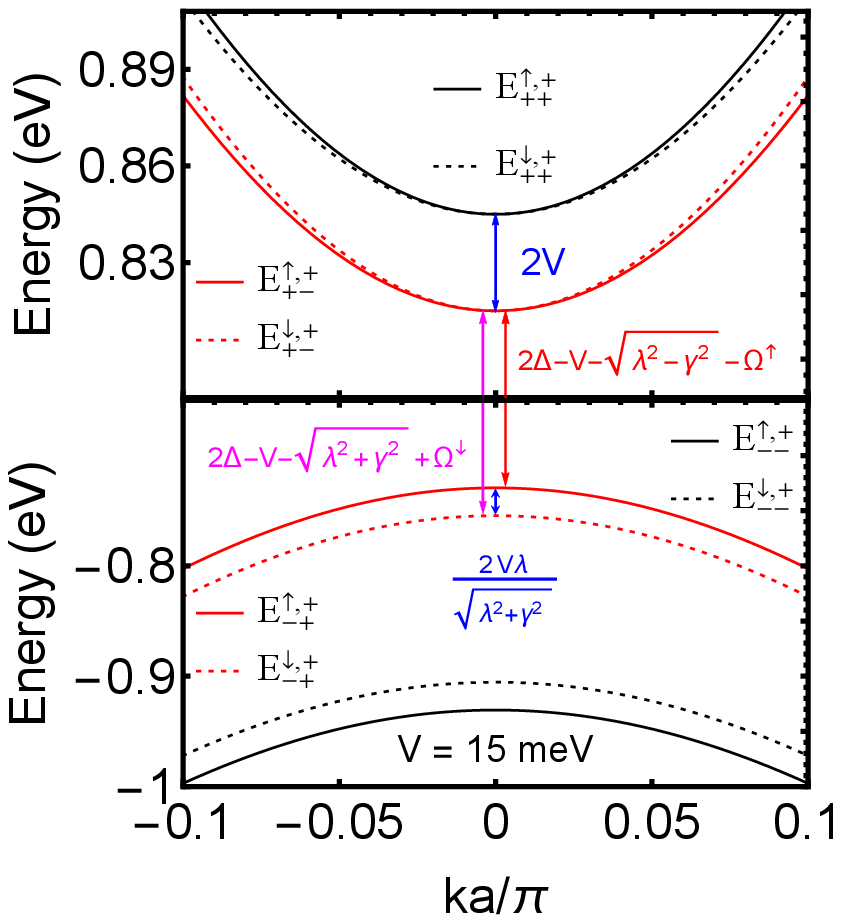}
\hspace*{0.77cm}
\includegraphics[width=.273\textwidth]{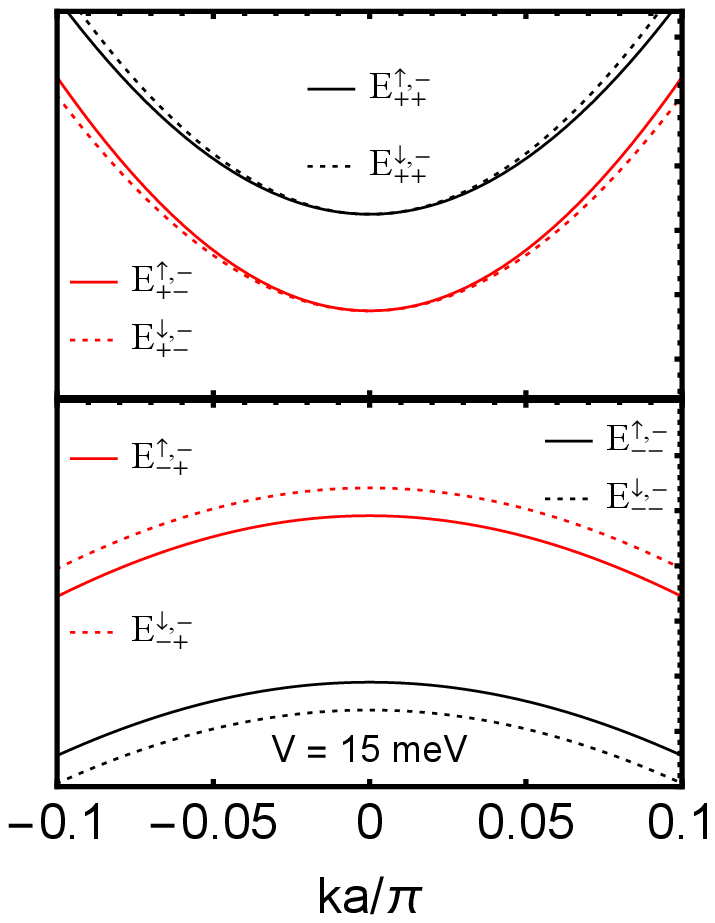}
\label{f1}
\vspace{-0.1cm}
\caption{Band structure of bilayer MoS$_{2}$ for $\lambda=0.074$ eV and $\gamma=0.047$ eV. The upper panels are for zero electric field energy ($V=0$) and the lower ones for $V=15$ meV. The left (right) panels are for the $K$ ($K^{\prime}$) valley and $\Omega^{s}=s\lambda V/[\lambda^{2}+\gamma^{2}]^{1/2}$.}
\end{figure}

\section{Formulation and electronic spectrum}
The one-electron Hamiltonian of bilayer MoS$_{2}$ near the $K$ and $K^{\prime}$ valleys \cite{9,10,25, 7i} reads 
%
\begin{equation}
H^{\tau}=
\begin{pmatrix}
-\xi_{1}^{s \tau} && v_{F}\pi_{-}^{\tau} && \gamma && 0\\
v_{F}\pi_{+}^{\tau} && \xi_{2}^{s \tau} && 0 && 0\\
\gamma && 0 && -\xi_{3}^{s \tau} && v_{F}\pi_{+}^{\tau}\\
0 && 0 && v_{F}\pi_{-}^{\tau} && \xi_{4}^{s \tau}
\end{pmatrix}. \label{e1}
\end{equation}
Here, $\tau=1 (-1)$ is for $K$ ($K^{\prime}$) valley, $\pi_{\pm}^{\tau}=\tau\pi_{x}\pm i\pi_{y}$, $\xi_{1}^{s \tau}=\kappa+\tau s \lambda+s M_{z}-\tau M_{v}$, $\xi_{2}^{s \tau}=\alpha-s M_{z}+\tau M_{v}$, $\xi_{3}^{s \tau}=\alpha-\tau s \lambda-s M_{z}+\tau M_{v}$, $\xi_{4}^{s \tau}=\kappa+s M_{z}-\tau M_{v}$ with $\kappa=\Delta+V$ and $\alpha=\Delta-V$ with $\Delta$  the monolayer band gap. Further, $v_{F}= $0.53$\times$10$^{6}$ m/s \cite{4} is the Fermi velocity, $V$   the external electric field energy, $\lambda$   the strength of the intrinsic SOC with spins up (down) represented by $s=+1(\uparrow)(s=-1(\downarrow))$, 
and $\gamma$  the effective interlayer interaction energy. Moreover, $M_{z}=g^{\prime}\mu_{B}B/2$ is the Zeeman exchange field induced by ferromagnetic order, $g^{\prime}$ the Land\'{e} $g$ factor $(g^{\prime}=g_{e}^{\prime}+g_{s}^{\prime})$, and $\mu_{B}$ the Bohr magneton \cite{z1}; $g_{e}^{\prime}=2$ is the free electron $g$ factor and $g_{s}^{\prime}=0.21$ the out-of-plane factor due to the strong SOC in MoS$_{2}$. The term, $M_{v}=g_{v}^{\prime} \mu_{B} B/2$ breaks the valley symmetry of the levels and $g_{v}^{\prime}=3.57$ \cite{z1}. The 
  valley splitting has been measured in very recent experiments \cite{z2, z3, z4, z5} and is theoretically shown to be approximately $30$ meV by first-principles calculations \cite{z6}.  The eigenvalues $E_{\mu}^{s,\tau}(k)$  
of Eq. (1), when the magnetic field is absent, are
\begin{equation}
E_{\mu}^{s,\tau}(k)=\hslash v_{F} \varepsilon_{\mu}^{s,\tau}(k).\label{e2}
\end{equation}
The subscript 
$\mu=(\mu_{1},\mu_{2})$ is used to denote the positive and  negative energies of the upper layer, by $\mu_{1}=\pm1$, and of the lower layer by $\mu_{2}=\pm1$. 
The factor $\varepsilon_{\mu}^{s,\tau}(k)\equiv\varepsilon$ in Eq. (\ref{e2}) is  the solution of the 
fourth-degree 
equation 
\begin{equation}
\left[ \left(\varepsilon-\alpha^{\prime}\right) \left( \varepsilon + \kappa^{\prime}-\tau s\lambda^{\prime}\right)-k^{2} \right] \left[ \left(\varepsilon-\kappa^{\prime}\right) \left( \varepsilon+ \alpha^{\prime}+ \tau s\lambda^{\prime}\right)-k^{2} \right] -\gamma^{\prime 2} \left( \varepsilon-\alpha^{\prime}\right)\left( \varepsilon-\kappa^{\prime}\right)=0,\label{e3}
\end{equation}
where $k\equiv k_y$ is the wave vector, $\varepsilon=E/\hslash v_{F}$, $\lambda^{\prime}=\lambda/\hslash v_{F}$, $\kappa^{\prime}=\kappa/\hslash v_{F}$, $\gamma^{\prime}=\gamma/\hslash v_{F}$, and $\alpha^{\prime}=\alpha/\hslash v_{F}$. In the combined limit   $\lambda^{\prime}\rightarrow 0$, $\kappa^{\prime}\rightarrow 0$, $\alpha^{\prime}\rightarrow 0$, we obtain the  energy dispersion for bilayer graphene  \cite{26}.

 In the upper panels of Fig. 1  we plot the energy dispersion of bilayer  MoS$_{2}$ for field $E_z=0$ ($V=0$ meV) at both valleys. We remark the following: (i) The splitting due to the SOC is zero in the conduction and valence bands even in the presence of SOC \cite{6, 7, 8, 9, 10, 7i, 25}. (ii) The splitting due to  interlayer hopping is zero in the conduction band but finite in the valence band \cite{6, 7, 8, 9, 10, 7i, 25}. Further, the splitting in the valence band is a combined effect of inter-layer coupling and SOC given by $2[\lambda^{2}+\gamma^{2}]^{1/2}$ at $k=0$. 
 This relation indicates that the valence band is still split  for $\lambda=0$ \cite{25}. (iii) The 
gap between conduction and valence band edges is given by $2\Delta-[\lambda^{2}+\gamma^{2}]^{1/2}$ for $k=0$ \cite{25}. 
 Notice that the effects of SOC and interlayer coupling are negligible in the conduction band, near $k=0$,  while at large values of $k$ the  SOC   effect 
 dominates.  
\begin{figure}[t]
\includegraphics[width=.35\textwidth]{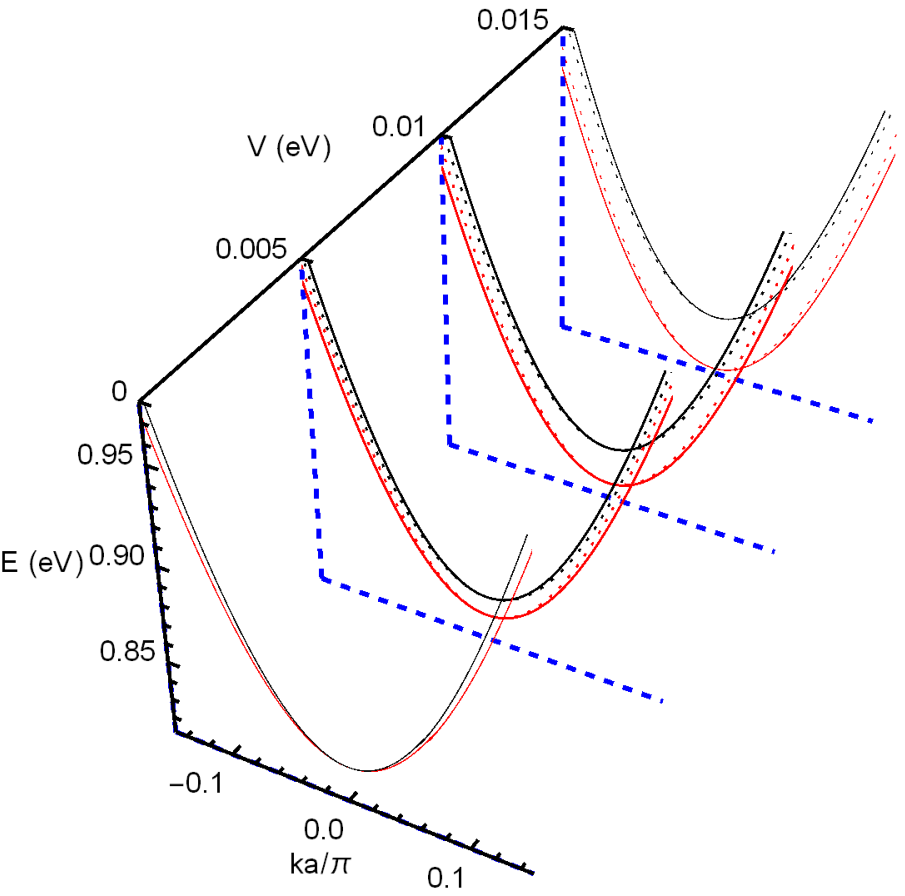}
\includegraphics[width=.35\textwidth]{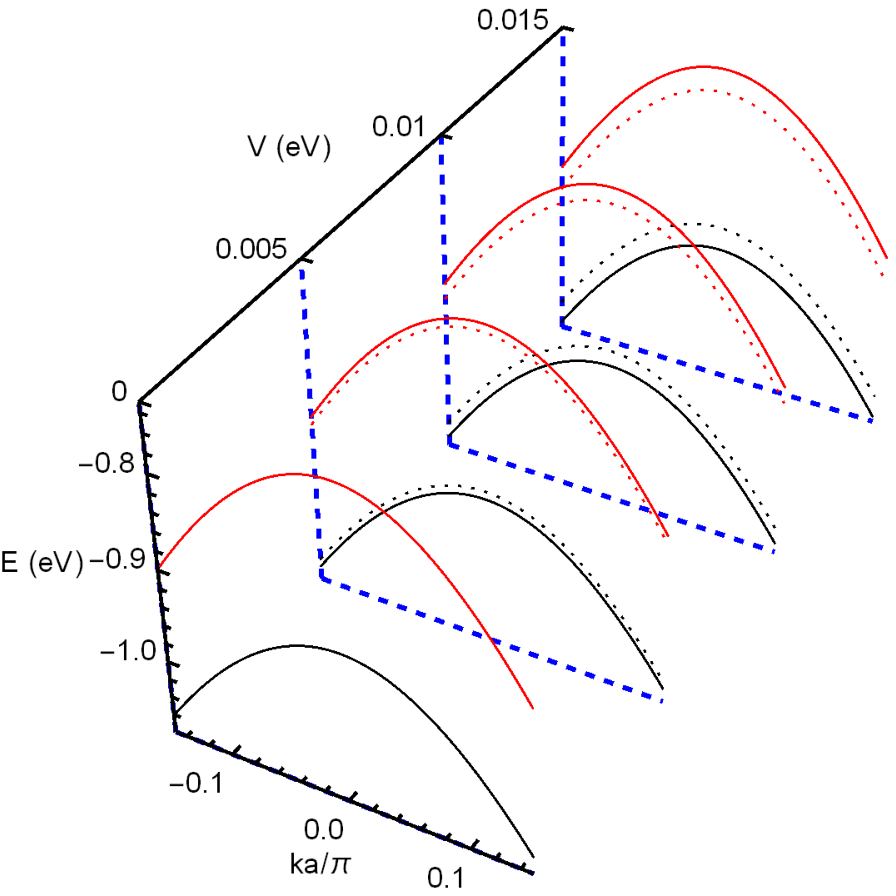}
\label{f2}
\vspace{-0.5cm}
\caption{Band structure of bilayer MoS$_{2}$ for different electric field $E_z$. The left (right) panel is for the conduction  (valence) band. The curve marking and  parameters are as 
Fig. 1.}
\end{figure}
 
 	For a finite 
	field $E_z$ 
	($V=15$ meV) we plot the energy spectrum in the lower panels of Fig. 1. We remark the following: (i) The SOC splitting is  modified by the field $E_z$. 
 We also note that the spin splitting in the conduction band due to the SOC is negligible for  the parameters and scale used.
 On the other hand, the valence band completely dictates the lifting of the spin degeneracy. (ii) An interlayer splitting is obtained in both the conduction and  valence bands. Analytically we obtain the gaps 
$2V\lambda/[\lambda^{2}+\gamma^{2}]^{1/2}$, for $V\ll\lambda$, and $2V$ at the valence  and conduction band edges, respectively. (iii) The band gap is also reduced by the  field $E_z\propto V$. It is equal to 
$2\Delta-V-[\lambda^{2}+\gamma^{2}]^{1/2}- \tau s\lambda V/[\lambda^{2}+\gamma^{2}]^{1/2}$ for $V\ll\lambda$. 
The spin and layer splittings
 increase with the field $E_z$ \cite{7, 8, z7}  or energy $V$, which can  be seen in Fig. (2).
 So far we assumed that the band edges are
 at the K point of the Brilloiun zone but this may not be the case neither for the valence band
 nor for the conduction band. In fact,  there are  arguments that our assumption holds 
\cite{2i, 9, 10, new1, new2} but DFT
calculations and a recent ARPES measurement \cite{new3} 
indicate that  the valence 
 band edge is shifted to the $\Gamma$ point.

%
\subsection{Landau levels}

In the presence of a magnetic field $B$ perpendicular  to the layers we replace $\mathbf{\pi}$ by $-i\hslash\mathbf{\nabla}+\mathbf{A}$ in Eq. (\ref{e1}) and take the vector potential $\mathbf{A}$  in the Landau gauge $\mathbf{A}=(0,Bx,0)$. 
After diagonalizing Eq. (\ref{e1}) the LL spectrum is obtained as  
\begin{equation}
E_{n,\mu}^{s, \tau}=
\hslash \omega_{c}
\,\varepsilon_{n,\mu}^{s, \tau},\label{e4}
\end{equation}
where $\omega_{c}=v_{F}\sqrt{2eB/\hslash}$ is the cyclotron frequency. The subscript $\mu=(\mu_{1},\mu_{2})$ is used to denote the positive and  negative energies  in the upper $(\mu_{1}=\pm1)$ and lower  $(\mu_{2}=\pm1)$ layers. For $n\geq1$ the factor $\varepsilon_{n,\mu}^{s, \tau}\equiv\varepsilon$ 
is  the solution of the fourth-order equation
\begin{equation}
\left[ \left(\varepsilon+d_{1}^{s \tau}\right) \left( \varepsilon - d_{2}^{s \tau}\right)-n \right] \left[ \left(\varepsilon+d_{3}^{s \tau}\right) \left( \varepsilon- d_{4}^{s \tau}\right)-(n+1) \right] - t^{2} \left( \varepsilon-d_{2}^{s \tau}\right)\left( \varepsilon- d_{4}^{s \tau}\right)=0,\label{e5}
\end{equation}
%
%
where $t = \gamma /\hslash \omega_{c}$, $d_{1}^{s \tau}=\kappa^{\tau}+s\lambda+\tau(sM_{z}-\tau M_{v})/\hslash \omega_{c}$, $d_{2}^{s \tau}=\alpha^{\tau}-\tau(sM_{z}-\tau M_{v}) /\hslash \omega_{c}$, $d_{3}^{s \tau}=\alpha^{\tau}-s\lambda-\tau(sM_{z}-\tau M_{v}) /\hslash \omega_{c}$, and $d_{4}^{s \tau}=\kappa^{\tau}+\tau(sM_{z}-\tau M_{v}) /\hslash \omega_{c}$ with $\kappa^{\tau}=\Delta+\tau V$ and $\alpha^{\tau}=\Delta-\tau V$ are dimensionless parameters. The eigenfunctions are 
\begin{equation}
\psi_{n,\mu}^{s,+}=
\frac{1}{\sqrt{L_{y}}}
\begin{pmatrix}
\varrho_{n,\mu}^{s,+} \phi_{n}\\
\Theta_{n,\mu}^{s,+}\, \phi_{n-1}\\
\Lambda_{n,\mu}^{s,+}\,\phi_{n}\\
\Upsilon_{n,\mu}^{s,+}\,\phi_{n+1}
\end{pmatrix}
e^{ik_{y}y}\,, \quad \quad %
\label{e7}
%
\psi_{n,\mu}^{s,-}=
\frac{1}{\sqrt{L_{y}}}
\begin{pmatrix}
\Lambda_{n,\mu}^{s,-}\,\phi_{n}\\
\Upsilon_{n,\mu}^{s,-}\,\phi_{n+1}\\
\varrho_{n,\mu}^{s,-}\, \phi_{n}\\
\Theta_{n,\mu}^{s,-}\, \phi_{n-1}
\end{pmatrix}
e^{ik_{y}y}.
\end{equation}
The coefficients are given by $\Theta_{n,\mu}^{s,\tau}=
\sqrt{n}\,\varrho_{n,\mu}^{s,\tau}/[\varepsilon_{n,\mu}^{s,\tau}-d_{2}^{s \tau}] $, $\Lambda_{n,\mu}^{s,\tau}=k_{n,\mu}^{s,\tau}\varrho_{n,\mu}^{s,\tau}$, and $\Upsilon_{n,\mu}^{s,\tau}=
\sqrt{n+1}\,k_{n,\mu}^{s,\tau}\,\varrho_{n,\mu}^{s,\tau}/[\varepsilon_{n,\mu}^{s,\tau}-d_{4}^{s \tau}]$,
with $\varrho_{n,\mu}^{s,\tau}$ 
the normalization constants
\begin{equation}
\varrho_{n,\mu}^{s, \tau}= \big\{(k_{n,\mu}^{s, \tau})^{2}\big[1+
(n+1)/(\varepsilon_{n,\mu}^{s, \tau}-d_{4}^{s  \tau})^{2})\big]+1+
n/(\varepsilon_{n,\mu}^{s, \tau}-d_{2}^{s \tau})^{2}\big\}^{-1/2}\label{e8}
\end{equation} 
%
%
and $k_{n,\mu}^{s, \tau}=[(\varepsilon_{n,\mu}^{s, \tau}+d_{1}^{s \tau})(\varepsilon_{n,\mu}^{s, \tau}-d_{2}^{s  \tau})-n]/t(\varepsilon_{n,\mu}^{s,\tau}-d_{2}^{s  \tau})$. 
Therefore, the wave function of bilayer MoS$_{2}$ is a mixture of  Landau wave functions with indices $n-1$, $n$, and $n+1$. 

In Eq.~ (\ref{e7}) 
the index $n$ can take the  values: $n=-1,0,1,.....$. If $n$ or  $n\pm 1$ is 
negative  the function $\phi_n$ or $\phi_{n\pm 1}$ is identically zero, i.e., $\phi_{-2}\equiv0$ and $\phi_{-1}\equiv0$. 
For $n=-1$ Eq.~ (\ref{e7}) is just $\psi_{-1}^{s,+}=(0,0,0,\phi_{0})$ and $\psi_{-1}^{s,-}=(0,\phi_{0},0,0)$, i.e., $\varrho_{n,\mu}^{s,\pm}$, 
$\Theta_{n,\mu}^{s,\pm}$, 
and $\Lambda_{n,\mu}^{s,\pm}$ 
are equal to zero. There is only one energy level per valley corresponding to $n=-1$. For $n=0$, Eq.~ (\ref{e7}) has zero coefficients $\Theta_{n,\mu}^{s,+}$ and $\Theta_{n,\mu}^{s,-}$, which results in three energy levels for each valley. For other values of $n$, i.e., for $n\geq1$, there are four eigenvalues of the Hamiltonian (\ref{e1}), corresponding to four Landau levels in a bilayer for a given valley $\tau=\pm1$.

In addition, there are two special LLs of bilayer MoS$_{2}$. 
  For  $n=-1$ and $n=0$, Eq. (\ref{e1}) takes, respectively, the forms 
\begin{equation}
H_{n=-1}^{+}=\xi_{4}^{+} \,, \quad \quad 
H_{n=-1}^{-}=\xi_{2}^{-}
\label{e-1}
\end{equation}
and
\begin{equation}
H_{n=0}^{+}=
\begin{pmatrix}
-\xi_{1}^{s +} && \gamma && 0\\
\gamma && -\xi_{3}^{s +} && \hslash \omega_{c} \\
0 && \hslash \omega_{c}  && \xi_{4}^{s +}
\end{pmatrix}\,, \quad \quad 
\label{e0} 
H_{n=0}^{-}=
\begin{pmatrix}
-\xi_{1}^{s -} && \hslash \omega_{c} && \gamma\\
\hslash \omega_{c} && \xi_{2}^{s -} && 0 \\
\gamma && 0  && -\xi_{3}^{s -}
\end{pmatrix}.
\end{equation}

\begin{figure}[t]
\centering
\includegraphics[width=.49\textwidth]{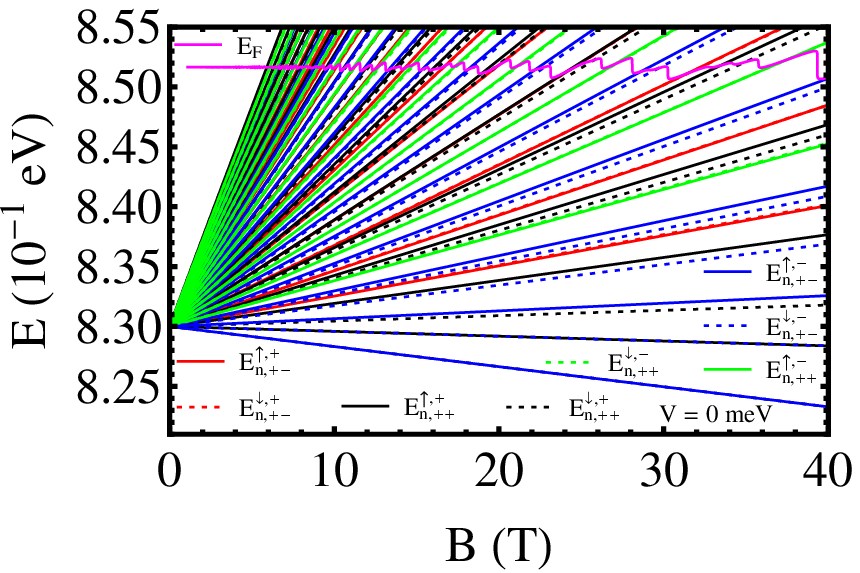}
\includegraphics[width=.49\textwidth]{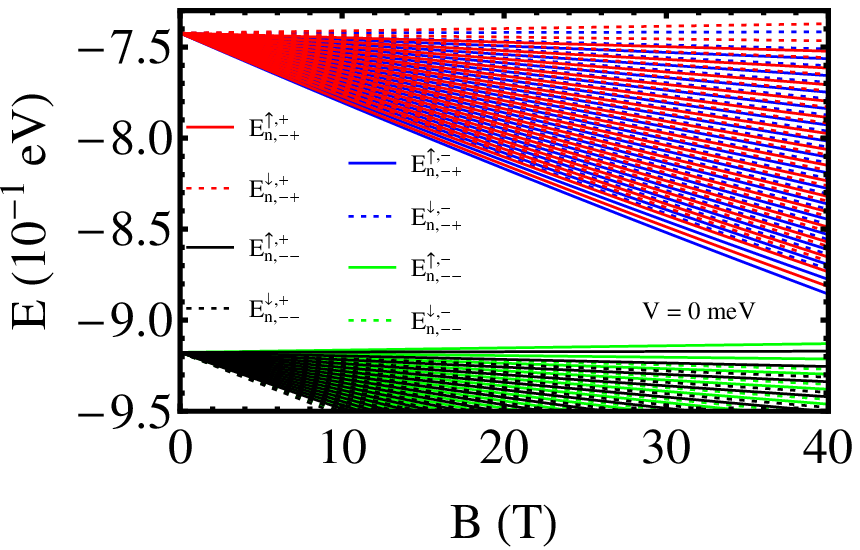}
\vspace{-0.5cm}
\caption{Energy spectrum of bilayer MoS$_{2}$ versus magnetic field $B$ for $M_{z}, M_{v}\neq 0$  and V=0. The left (right) panel is for the conduction     Ê(valence) band. The magenta curve shows the Fermi energy $E_F$ versus $B$ for an electron density $n_{e}= 1.9\times 10^{13}$ cm$^{-2}$.} 
\label{f3}
\end{figure}

The factor $\varepsilon$ corresponding to Eq. (\ref{e0}) is given by the roots of the cubic equation
\begin{align}
&\left( \varepsilon+ d_{1}^{s \tau}\right) \left[ \left(\varepsilon+d_{3}^{s  \tau}\right) \left( \varepsilon- d_{4}^{s  \tau}\right)-1   \right] -t^{2} \left( \varepsilon-d_{4}^{s  \tau}\right)=0.\label{e10}
\end{align}
%
%
The corresponding eigenstates take the form 
\begin{align}
&\psi_{0,\mu}^{s,+}=
\frac{1}{\sqrt{L_{y}}}
\begin{pmatrix}
\varrho_{0,\mu}^{s,+}\, \phi_{0}\\
\,0\\
\Lambda_{0,\mu}^{s,+}\,\phi_{0}\\ 
\Upsilon_{0,\mu}^{s,+}\,\phi_{1}
\end{pmatrix}
e^{ik_{y}y}\,, \quad \quad %
\label{e12}
&\psi_{0,\mu}^{s,-}=
\frac{1}{\sqrt{L_{y}}}
\begin{pmatrix}
\Lambda_{0,\mu}^{s,-}\,\phi_{0}\\ 
\Upsilon_{0,\mu}^{s,-}\,\phi_{1}\\
\varrho_{0,\mu}^{s,-}\,\phi_{0}
\\ 0
\end{pmatrix}
e^{ik_{y}y}\,.
\end{align}
Note that Eqs.~(\ref{e10}) 
give only three roots while $\mu$ provides four labels. We reserve the labels $\mu=(+, +)$ for the fourth root and denote by 
$\varepsilon_{-1,++}^{s,+}=d_{4}^{s +}$ the corresponding eigenvalue for $n=-1$. 
We then write  the respective LL state as $\psi_{-1,++}^{s,+}=(0,0,0,\phi_{0})^{T}e^{ik_{y}y}/\sqrt{L_{y}}$, where $T$ denotes the transpose of the row vector. 
Further, we reserve the label $\mu=(+,-)$ for $n=-1$  at the $K^{\prime}$ valley irrespective of the $K$ valley, since the corresponding eigenvalue is $\varepsilon_{-1,+-}^{s,-}=d_{2}^{s -}$ and yields the  state  $\psi_{-1,+-}^{s,-}=(0,\phi_{0},0,0)^{T}e^{ik_{y}y}/\sqrt{L_{y}}$. 
The eigenfunctions depend on the quantum numbers $n$ and $k_{y}$ but the eigenvalues are independent of $k_{y}$.

\subsection{Limiting cases}
(i) Setting $\gamma=V=0$ and $M_{z}=M_{v}=0$ in Eq. (\ref{e4})  gives the eigenvalues of a MoS$_{2}$ monolayer  or two uncoupled and unbiased layers 
\begin{equation}
\varepsilon=- s\lambda_{1}\pm\big[(\Delta^{\prime}+ s\lambda_{1})^{2}+(n+1)\big]^{1/2}\,, \quad \quad 
%
\varepsilon=s\lambda_{1}\pm\big[(\Delta^{\prime}- s\lambda_{1})^{2}+n\big]^{1/2},\label{e13}
\end{equation}
where $\Delta^{\prime}=\Delta /\hslash \omega_{c}$ and $\lambda_{1}=\lambda /\hslash \omega_{c}$. These results are consistent with those in Refs. \onlinecite{5, 9i}. If we set $\Delta^{\prime}=\lambda_{1}=0$ in Eq. (\ref{e13}), we obtain the well-known eigenvalues for monolayer graphene \cite{9ii} 
\begin{equation}
\varepsilon=\pm\,\sqrt{n+1},\quad 
\varepsilon=\pm\,\sqrt{n}.\label{e14}
\end{equation}
(ii) For $\Delta=\lambda=V=M_{z}=M_{v}=0$, we obtain the 
LL spectrum of  bilayer graphene \cite{16, 17, 18}, 
\begin{equation} 
\varepsilon=\pm\dfrac{1}{
\sqrt{2}}
\Big[t^{ 2}+2(2n+1)
\pm 
\big[(t^{ 2}+2(2n+1))^{2}-16n(n+1)\big]^{1/2}\Big]^{1/2}. \label{e15}
\end{equation}  
This equation can be further simplified by expanding the internal square root in the limit $n\ll t^{ 2}$. 
Moreover, 
by taking the negative sign, the solution is
\begin{equation}
\varepsilon=\pm\ 
2\sqrt{n(n+1)}\big/t.\label{e16}
\end{equation}
%
\begin{figure}[t]
\centering
\includegraphics[width=.45\textwidth]{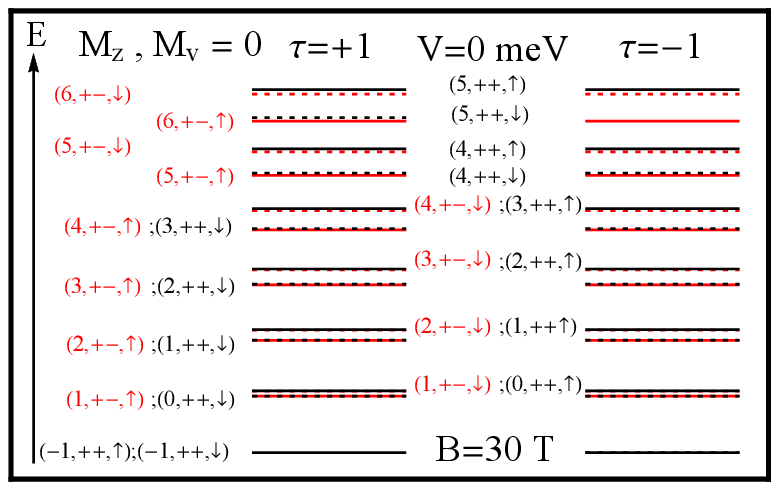}
\includegraphics[width=.45\textwidth]{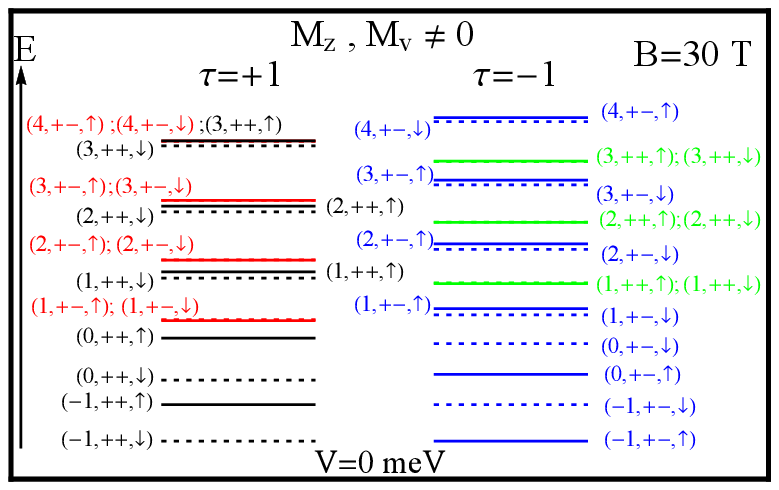}
\includegraphics[width=.45\textwidth]{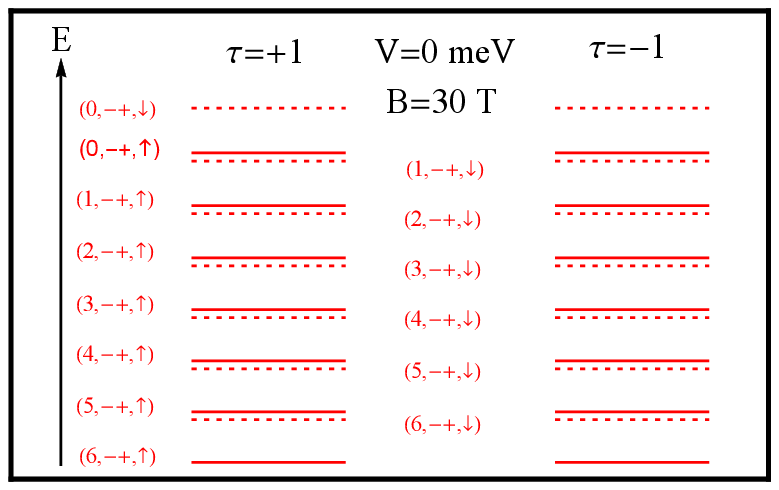}
\includegraphics[width=.45\textwidth]{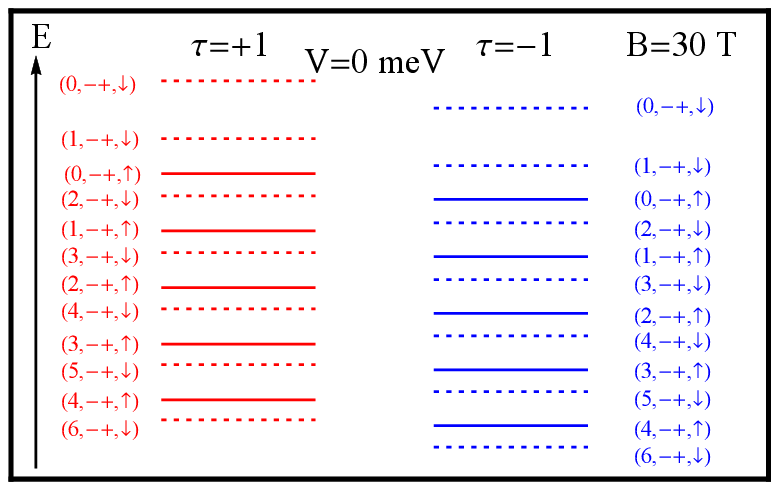}
\caption{LL spectrum of bilayer MoS$_{2}$ at $B=30$ T and $V=0$ labeled by $(n, \mu, s)$ with $s$ the spin index $s=\pm 1 (\uparrow\downarrow)$ and $\mu$ the layer index (see text after Eq.~ (\ref{e4})) $\mu=(\mu_{1} \mu_{2})$. The upper panels are for the conduction band and the lower ones are for the valence band. Further, the left panels are
for $M_{z}= M_{v}=0$  and the right ones  for  $M_{z}\neq M_{v}\neq 0$. For simplicity we do not show  the valence band levels for the second layer.}
\end{figure}
This spectrum  is similar to that of Refs. \onlinecite{16, 18} 
obtained by means of a $2\times2$ Hamiltonian. The energy of higher LLs is obtained by taking the $+$ 
sign in front of the internal square root in Eq. (\ref{e15}).
\begin{figure}[t]
\centering
\includegraphics[width=.47\textwidth]{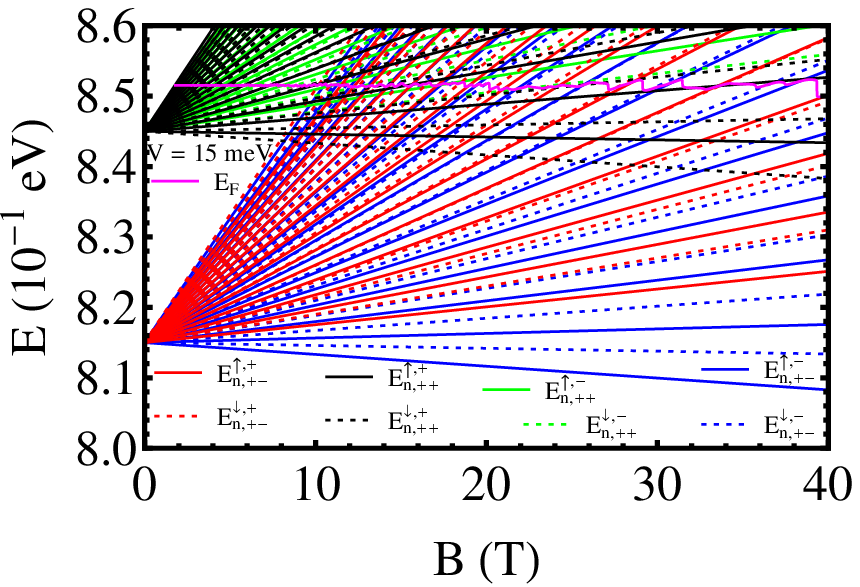}
\includegraphics[width=.49\textwidth]{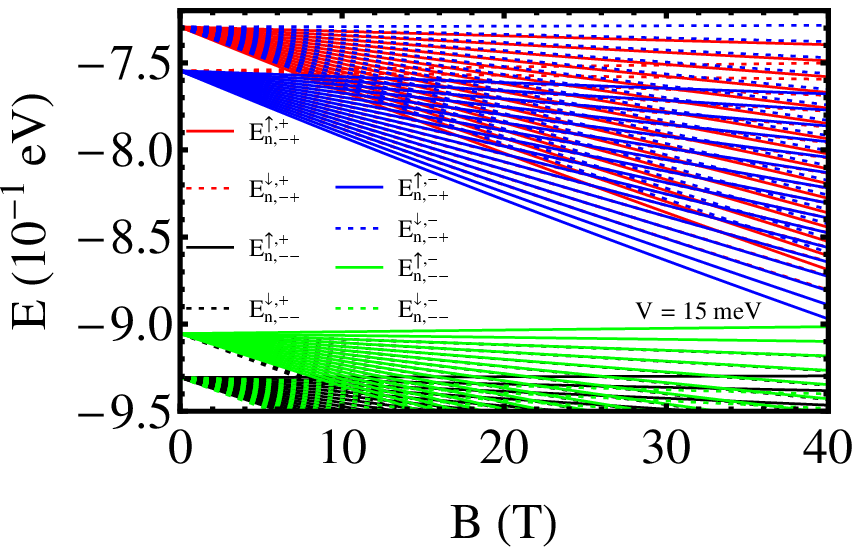}
\vspace{-0.5cm}
\caption{As in Fig. 3 but for $V= 15$ meV.}
\label{f4}
\end{figure}
%

In Fig. 3  we plot the spectrum given by Eq. (\ref{e4}) versus the  field $B$ for $V= 0$ and finite spin $M_{z}$ and 
$M_{v}$ Zeeman fields. The left panel is for the conduction band and the right one for the valence band. 
The main findings are as follows. (i) The energy spectrum 
grows linearly with the field $ B$ due to the huge band gap. (ii) For $B=0$ 
there are no LLs and the spin splitting in the conduction band, due to SOC, is very small \cite{6, 7, 8, 9, 10, 7i, 25}, as seen in the upper panels of Fig. 1. But for a finite  field $B$  we obtain a significant spin splitting: for $B=30$ T this is seen  in the left panels of Fig. 4 and is due to the SOC alone, expressed by the term $\tau s \lambda$ in Eq. (1), since we intentionally set $M_{z}=M_{v}=0$. The right panels in Fig. 4  are for $M_{z}\neq M_{v}\neq0$ Interestingly, the spin splitting energy increases with $ B$. 
Within  the same LL  $n=10$ in the conduction band it  is $1.4$ meV at $B=10$ T, $2.8$ meV at $B=20$ T, and $4.1$ meV at $B=30$ T. Further, one noteworthy feature is that the spin splitting among adjacent 
smaller-index LLs is unobservable, i.e. $E_{1,+-}^{\uparrow(\downarrow),+}\cong E_{0,++}^{\downarrow(\uparrow),+}$ and $E_{0,+-}^{\uparrow(\downarrow),-}\cong E_{1,++}^{\downarrow(\uparrow),-}$, whereas it 
is enhanced among the higher-index LLs due to the combined effect of the SOC and interlayer coupling terms in contrast with monolayer MoS$_{2}$ \cite{9i}. (iii) In the presence of the Zeeman fields 
the LL energies for spin up (down) at the $K$ valley are different 
than those with spin down (up)  at the $K^{\prime}$ valley and  lead to 
spin and valley polarizations contrary to the $B=0$ case in which they are the same 
\cite{6, 7, 8, 9, 10, 7i, 25}. (iv) For 
$M_{z}\neq 0$ and $M_{v}=0$  the spin splitting in the conduction band ($n=10$) is $1.9$ meV at $B=10$ T, $3.7$ meV at $B=20$ T and $5.4$ meV at $B=30$ T. (v) The spin splitting among the lower and upper layer LLs at the $K$ and $K^{\prime}$ valleys has vanished i.e. $E_{1,+-}^{\uparrow,+}\cong E_{1,+-}^{\downarrow,+}$ and $E_{2,++}^{\uparrow,-}\cong E_{2,++}^{\downarrow,-}$. This unexpected behaviour of LLs is due to the presence of the $M_{v}\neq 0$ term. We also notice  that the splitting is  unobservable between other LLs e.g. $E_{4,+-}^{\uparrow(\downarrow),+}\cong E_{3,++}^{\uparrow,+}$, $E_{8,++}^{\uparrow(\downarrow),+}\cong E_{9,+-}^{\uparrow(\downarrow),-}$, $E_{9,+-}^{\uparrow(\downarrow),+}\cong E_{8,++}^{\uparrow(\downarrow),-}$, $E_{14,+-}^{\uparrow(\downarrow),+}\cong E_{13,++}^{\uparrow(\downarrow),+}$ and $E_{14,+-}^{\uparrow(\downarrow),-}\cong E_{13,++}^{\uparrow(\downarrow),-}$. Such a behaviour of the LLs is absent in monolayer MoS$_{2}$\cite{9i}. However, the value of the spin splitting is very strong in the valence band for both valleys.    
(vi) For $M_{z}=M_{v}=0$, the $n=0$ level is two-fold spin-split and 
 valley degenerate in both the conduction and valence bands.
  For  finite Zeeman fields though 
 it is  spin and valley non-degenerate   in both the conduction and valence band. As for  the  $n=-1$ level, it is spin and valley degenerate for $M_{z}=M_{v}=0$ whereas it is spin non-degenerate and valley degenerate for $M_{z}\neq M_{v}\neq 0$ in the conduction band $(\Delta\pm sM_{z}-M_{v})$ with plus $(+)$ sign for the $K$ valley and negative $(-)$ sign for the $K^{\prime}$ valley; that is, the spin splitting is the same but opposite in both valleys. On the other hand, there is no level  in the valence band for $n=-1$. These distinct features of the $n=0$ and $n=-1$ levels can clearly be seen in  Fig. 4. 
 (vii) The LLs are unevenly spaced in the conduction band but equidistant in the valence band. This difference arises from  the lack of electron-hole symmetry in our system. This unusual behavior of the LLs can clearly be seen in Fig. 4 for both  zero and finite Zeeman fields.
\begin{figure}[t]
\centering
\includegraphics[width=.45\textwidth]{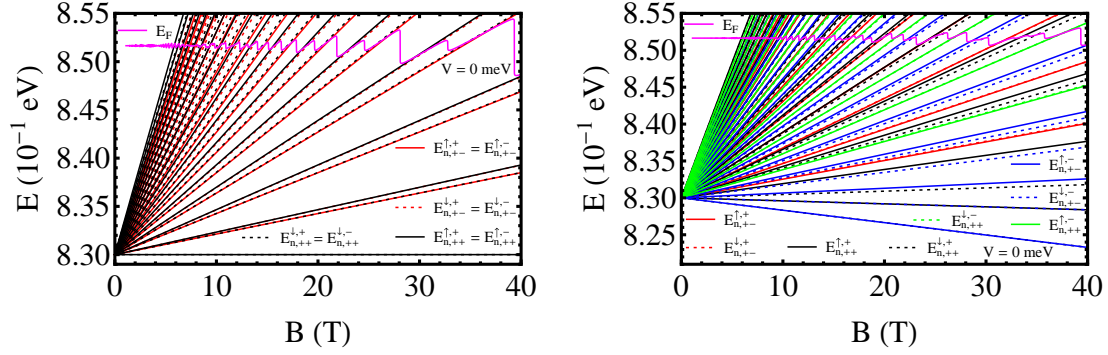}
\includegraphics[width=.45\textwidth]{l1.eps}
\vspace{-0.5cm}
\caption{LLs in bilayer MoS$_{2}$ (conduction band) 
vs the magnetic field $B$ 
for $V=0$ meV. 
The left panel is for $M_{z}=M_{v}=0$, the right one for $M_{z}\neq M_{v}\neq0$. The magenta curve shows $E_{F}$ vs $B$.} 
\end{figure}

We show the LL spectrum in Fig. 5 for finite  field $E_z$ ($V=15$ meV) including the $M_{z}$ and $M_{v}$ terms. We deduce the following: (i) The field $E_z$ modifies 
the inter-layer splitting, e.g., it makes it $30$ meV in the conduction band. (ii) For $M_{z}=M_{v}=0$ the LLs are still doubly degenerate consisting of a spin-up ($\uparrow$) state from the $K$ valley and a spin-down ($\downarrow$) state from the $K^{\prime}$ valley. Furthermore, the $n=0$ LL is spin non-degenerate but valley-degenerate in the conduction band. However, its  spin and valley degeneracy 
 are completely lifted in the valence band. Moreover, the valley degeneracy of the $n=-1$ level is lifted while its spin degeneracy in the conduction band is not. 
 Interestingly, the 
spin splitting energy between  adjacent LLs is also enhanced due to the finite  field $E_z$. For example, for $n=10$ its value is $1.9$ meV at $B=10$ T, $3.7$ meV at $B=20$ T, and $5.3$ meV for $B=30$ T. (iii) For $M_{z}\neq 0$ and $M_{v}=0$ the spin splitting in the conduction band, for $n=10$, is $3.2$ meV at $B=10$ T, $6.2$ meV at $B=20$ T, and $9.1$ meV at $B=30$ T. Additionally, the spin and valley degeneracies of all LLs in the conduction and valence bands are lifted. (iv) The energies of the LLs for the lower and upper layers  have 
different slopes in $B$ 
leading to level crossings. Interestingly, these crossings give rise to  additional degeneracies of the levels. From Eq. (5) with $t=0$, we obtain  that these  degeneracies, 
at specific energies and fields,  are embodied in the 
relation
\begin{figure}[t]
\centering
\includegraphics[width=.45\textwidth]{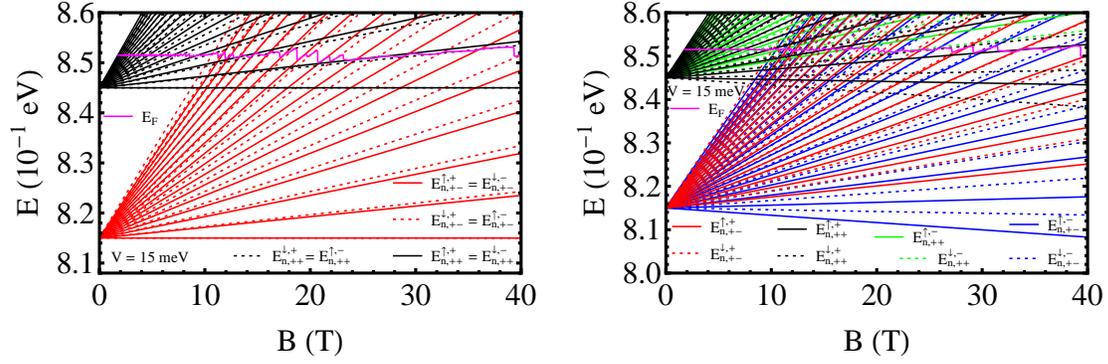}
\includegraphics[width=.45\textwidth]{l7.eps}
\vspace{-0.5cm}
\caption{As in Fig. 5 but for $V=15$ meV.}
\end{figure}
\begin{equation}
n_{1}+n_{2}=\varepsilon_{n_{1}}^{2}+\varepsilon_{n_{2}}^{2}+\varepsilon_{n_{1}}(d_{1}^{s\tau}-d_{2}^{s\tau})+\varepsilon_{n_{2}}(d_{3}^{s\tau}-d_{4}^{s\tau})-(d_{1}^{s\tau}d_{2}^{s\tau}+d_{3}^{s\tau}d_{4}^{s\tau})-1.
\end{equation}
Here $n_{1}$ and $n_{2}$ indices label the LLs in  the lower and upper layers, respectively. For $ \Delta = \lambda=M_{z}=M_{v} = 0 $ we obtain a relation similar to that  in  unbiased bilayer graphene \cite{16}.  Also, though not shown, for $V\neq 0$ 
the LL spacing is not uniform in the conduction band whereas it is  in the valence band and the spectra are similar  to those in Fig. 4.

%

 The Fermi energy $E_F$, at constant electron concentration $n_{e}$, is obtained from the relation 
\begin{equation}
n_{e}= \int_{-\infty}^{\infty} D(E)f(E) dE=\dfrac{g_{s\slash v}}{D_{0}}\sum_{n,\tau,s,\mu}f(E_{n,\mu}^{s,  \tau}),\label{e18}
\end{equation}
where $f(E_{n,\mu}^{s, \tau})=1/\big[1+\exp[\beta(E_{n,\mu}^{s, \tau}-E_{F})]\big]$,  $\beta=1/k_{B}T$, is the Fermi-Dirac 
function,  $D(E)$  the density of states (DOS),  and
$D_{0}=2\pi l_{B}^{2}$; $g_{s} (g_v)$ denotes the spin (valley) degeneracy.
\begin{figure}[t]
\centering
\includegraphics[width=.47\textwidth]{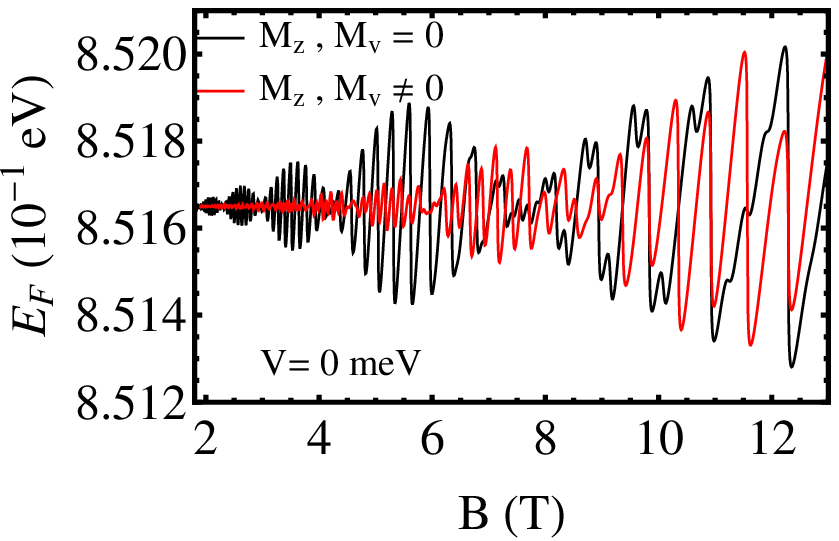}
\includegraphics[width=.47\textwidth]{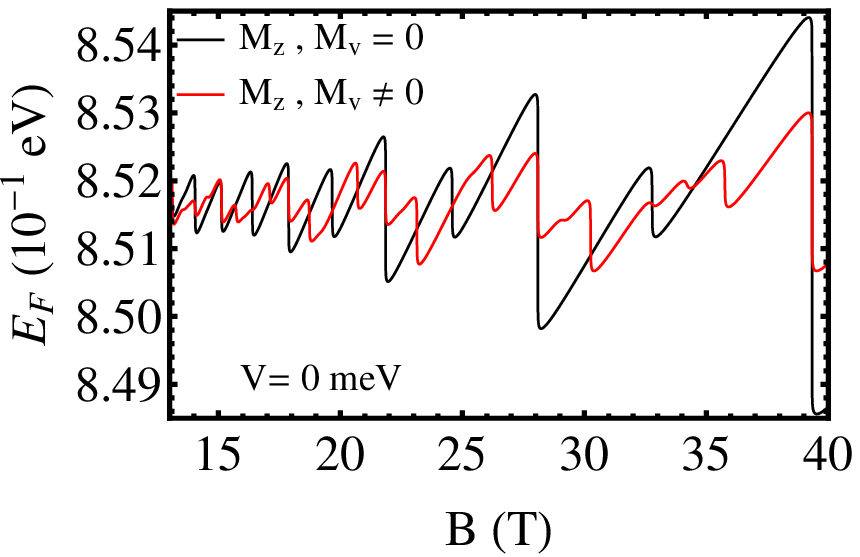}
\includegraphics[width=.47\textwidth]{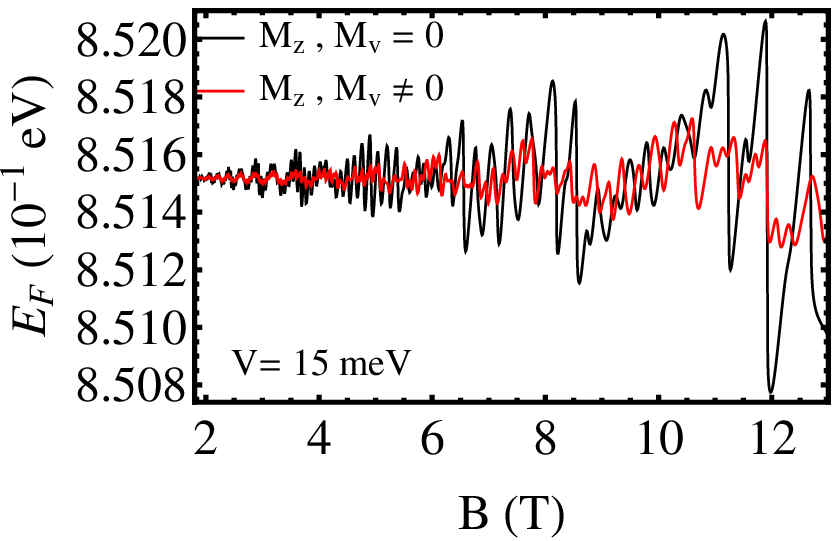}
\includegraphics[width=.47\textwidth]{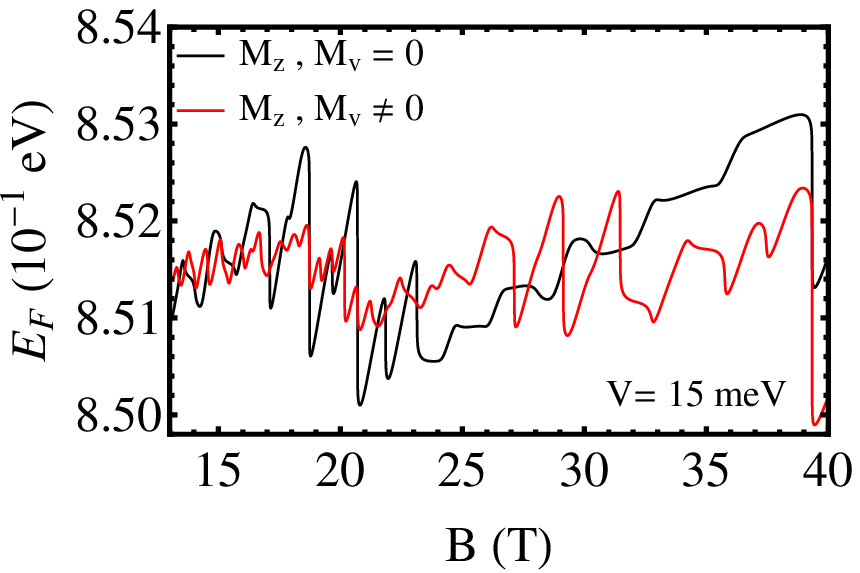}
\vspace{-0.5cm}
\caption{Fermi energy $E_F$ versus $B$ at $T=1$ K. The upper panels are for $V=0$ meV and lower ones for $V=15$ meV. The panels differ only in the  range  of $B$.}
\end{figure}

To  better  appreciate the  difference between zero and finite Zeeman fields we redraw,  in Fig. 6,  the LL spectrum in the left panel for $M_{z}=M_{v}=0$ and in the right one for $M_{z}\neq M_{v}\neq0$ as  functions of the magnetic field $B$. The LLs $(n\geq0)$ are spin non-degenerate and valley degenerate  for $M_{z}=M_{v}=0$ but for $M_{z}\neq M_{v}\neq0$ 
the valley degeneracy is lifted. Nevertheless,  the level for $n=-1$  is two-fold spin and valley-degenerate in the absence of the Zeeman terms {but its spin 
  degeneracy is lifted in their presence. }
The magenta solid lines in Fig. 6 show $E_F$ 
versus the  field $B$ for $V=0$ meV calculated numerically from Eq. (\ref{e18}). For zero Zeeman terms, the small intra-LL jumps indicate the presence of splitting due to SOC which is strengthened by the inter-layer coupling energy 
as seen in  the left panel of  Fig 6. However, the lifting of the spin and valley degeneracies due to finite Zeeman fields also give rise to additional intra LL small jumps in the $E_F$ curve as can be seen in  the right panel of  Fig 6.

 In Fig. 7 we replot the spectrum for $M_{z}=M_{v}=0$ and $M_{z}\neq M_{v}\neq 0$ at $V=15$ meV. We can see that the $n\geq 0$ levels are spin non-degenerate and valley degenerate for $M_{z}=M_{v}=0$ whereas they are spin and valley non-degenerate for $M_{z}\neq M_{v}\neq 0$. On the other hand,  the level for $n=-1$ is spin degenerate and valley non-degenerate in the absence of the Zeeman fields while its spin and valley degeneracies are lifted in their presence. 
For zero Zeeman fields (Fig. 7, left panel), the additional intra LL small jumps in the $E_F$ curve are due to the spin and inter-layer splittings which are modified by the  electric field $E_z$. However, the spin and valley non-degeneracies in the presence of the Zeeman fields lead to  additional intra-LL small jumps in $E_F$ as can be seen in the right panel of Fig. 7. 

In Fig. 8 we show $E_F$ as a function of the magnetic field for 
 $V=0$ meV and $V=15$ meV. $E_F$ shows not only the beating phenomenon at low fields $B\leq 13$ T but also dictates the giant splitting of the LLs at higher fields under the combined effect of spin and Zeeman terms as seen in the upper panels of Fig. 8. In the lower panels of Fig. 8 another worth noticing feature is the beating of the oscillations for $B$ fields up to about $8$ T with a giant splitting of the LLs at higher fields due the  field $E_z$ and the spin and valley Zeeman fields. 
\begin{figure}[t]
\centering
\includegraphics[width=.47\textwidth]{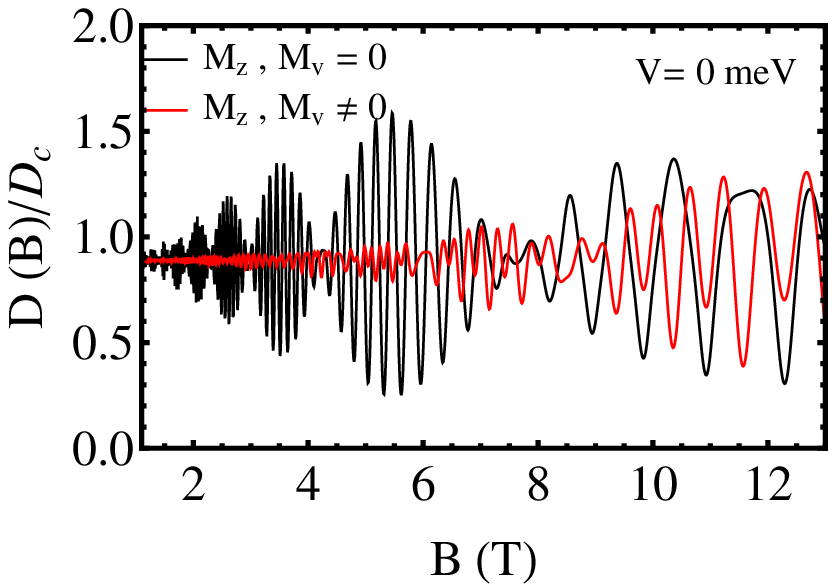}
\includegraphics[width=.47\textwidth]{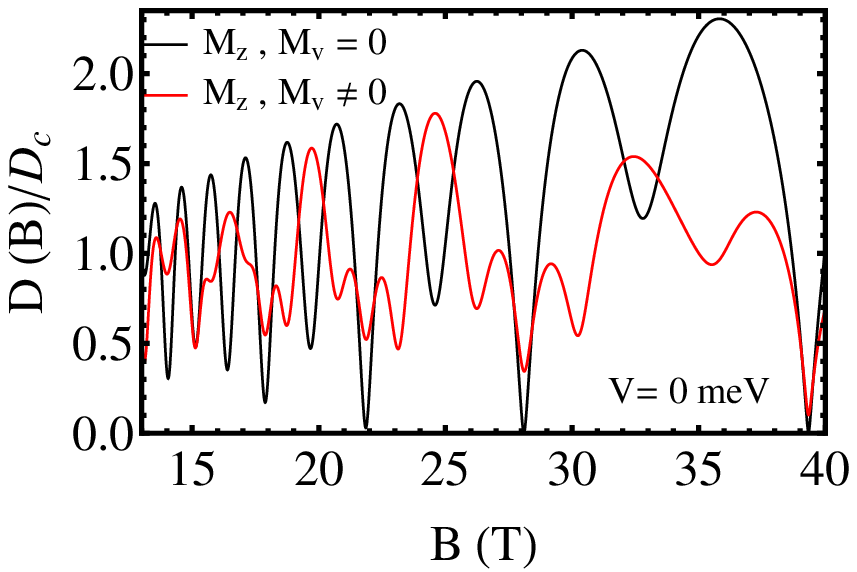}
\includegraphics[width=.47\textwidth]{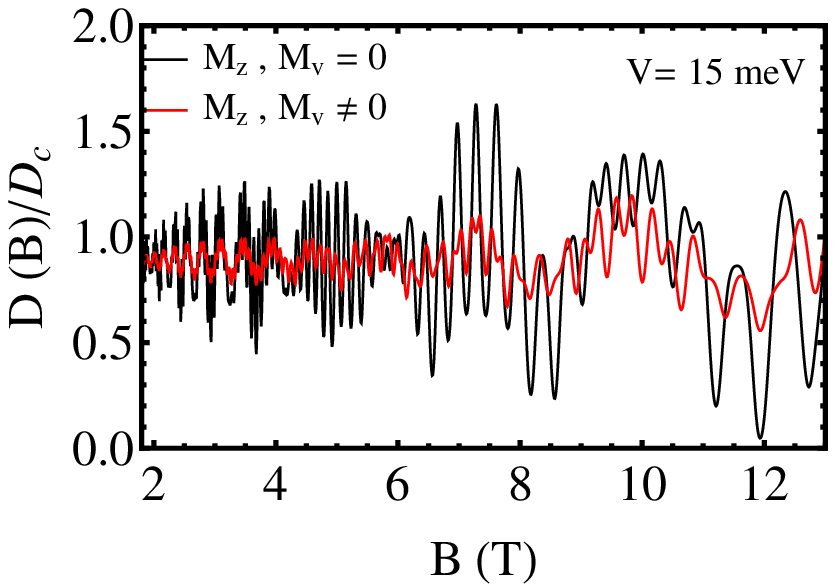}
\includegraphics[width=.47\textwidth]{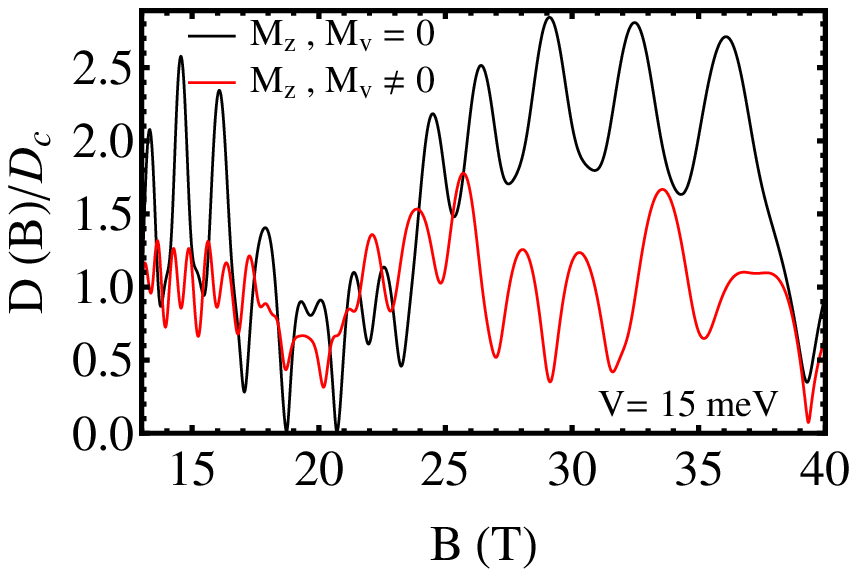}
\vspace{-0.5cm}
\caption{Dimensionless density of states (DOS) with $D_{c}=g_{s/v}/D_{0}\Gamma\sqrt{2\pi}$ vs $B$ for a LL width $\Gamma=0.1\sqrt{B}$ meV. The upper panels are for $V=0$ meV and the lower ones for $V=15$ meV. The left and right panels differ only in the magnetic field range ($x$ axis).}
\end{figure}

In Fig. 9 we plot the dimensionless DOS versus the  field $B$  in the conduction band for two different values of $E_z$. We observe a beating pattern at low  fields $B$ and a splitting at higher fields  in the SdH oscillations. The former and latter characteristics are due to the splitting of 
the LLs by the combined effect of the SOC,  interlayer coupling 
and Zeeman terms, and the layer splitting  modified by the field  $E_z$ as seen by contrasting the curves of the upper and lower panels. 
 One noteworthy feature is that the Zeeman fields and layer splitting suppress the amplitude of the beating at low $B$ fields and enhance the  oscillation amplitude 
at higher $B$ fields. At higher $B$ fields, the maximum SdH oscillation amplitude 
in the presence of the field $E_{z}$ occurs due to the LL degeneracy which  arises from the level crossings of the two layers. The inter-layer splitting and Zeeman effect  change the position and number of the beating nodes as compared to monolayer MoS$_{2}$ \citep{9i}.
 We notice that  in the conduction band the beating of the oscillations is observed in the range $0\leq B\leq13$ T, for $V=0$ meV, and in the range $0\leq B\leq 8$ T for $V=15$ meV. Above these ranges the beating pattern  is replaced by a split in the SdH oscillations. The particular beating oscillation pattern occurs when the level broadening is of the order of the cyclotron energy $\hslash \omega_c$ 
 and is replaced  by the split when the SOC  becomes weak at larger 
fields $B$.
\begin{widetext}

\section{Conductivities}
\subsection{Hall conductivity}

We use the linear-response theory as formulated in Ref. \onlinecite{24}. If one uses the identity 
$f_{\zeta}(1-f_{\zeta^{\prime}})[1-\exp(\beta(E_{\zeta}-E_{\zeta
^{\prime}}))]=\left(  f_{\zeta}-f_{\zeta^{\prime}}\right)  $, with $f_{\zeta}$ the Ferni-Dirac function, 
the Hall conductivity takes the simple form 
  \cite{9i, 17, 9ii, i2}, 
\begin{align}
&\sigma_{yx}=\frac{i\hslash e^{2}}{L_{x}L_{y}}\sum_{\zeta\neq\zeta^{^{\prime}
}}\frac{\left(  f_{\zeta}-f_{\zeta^{\prime}}\right)  \left\langle
\zeta\right\vert v_{x}\left\vert \zeta^{\prime}\right\rangle \left\langle
\zeta^{\prime}\right\vert v_{y}\left\vert \zeta\right\rangle }{\left(
E_{\zeta}-E_{\zeta^{\prime}}\right)  ^{2}},\label{e19}
\end{align}
with $\left\vert \zeta\right\rangle =\left\vert n,\mu,s,\tau,k_{y}\right\rangle $ and $\left\langle \zeta\right\vert v_{x}\left\vert\zeta^{\prime}\right\rangle $ and $\left\langle \zeta^{\prime}\right\vert
v_{y}\left\vert \zeta\right\rangle $  the off-diagonal matrix elements of the velocity operator. They are evaluated with the help of the corresponding operators $v_{x}=\partial H/\partial p_{x}$ and $v_{y}=\partial H/\partial p_{y},$ and are given in terms of the Pauli matrices $\sigma_{\upsilon}$
\begin{align}
\begin{split}
v_{x}=
\tau v_{F}
\begin{pmatrix}
\sigma_{x} && 0\\
0 && \sigma_{x}
\end{pmatrix}
,v_{y}=
v_{F}
\begin{pmatrix}
\sigma_{y} && 0\\
0 && -\sigma_{y}
\end{pmatrix},\label{e20}
\end{split}
\end{align}
%
With 
$\varepsilon_{n,d_{2}}\equiv\varepsilon_{n,\mu}^{s ,\tau}- d_{2}^{s \tau}$  , $\varepsilon_{n,d_{4}}\equiv\varepsilon_{n,\mu}^{s , \tau}- d_{4}^{s \tau}$
and $Q=v_{F}\varrho_{n,\mu}^{s , \tau} \varrho_{n^{\prime},\mu^{\prime}}^{s^{\prime} , \tau^{\prime}}\,\delta_{s,s^{\prime}}$ the results are
\begin{multline}
\hspace{-0.6cm}
\left\langle \zeta 
\right\vert v_{x}\left\vert \zeta^\prime 
\right\rangle   =\tau Q
  \Big[ \Big(    \frac{ \sqrt{n^{\prime}} 
}{  \ \varepsilon_{n, d_{2}}^\prime   }  +\frac{\sqrt{n+1}k_{n,\mu}^{s, \tau}\ k_{n^{\prime},\mu^{\prime}}^{s^{\prime}, \tau^{\prime}}}{\varepsilon_{n,d_{4}}  }     \Big)\delta_{n,n^{\prime}-1} 
+  \Big(    \frac{ \sqrt{n} 
}{  \ \varepsilon_{n,d_{2}}   }  +\frac{\sqrt{n^{\prime}+1}k_{n,\mu}^{s, \tau}\ k_{n^{\prime},\mu^{\prime}}^{s^{\prime},\tau^{\prime}}}{\varepsilon_{n,d_{4}}^{\prime}  }     \Big)\delta_{n,n^{\prime}+1} \Big],\label{e21}
\end{multline}
\begin{multline}
\hspace{-0.6cm}
\left\langle  \zeta^\prime 
\right\vert
v_{y}\left\vert  \zeta
\right\rangle   
=\tau iQ
  \Big[ \Big(    \frac{ \sqrt{n^{\prime}} 
}{  \ \varepsilon_{n,d_{2}}^\prime   }  +\frac{\sqrt{n+1}k_{n,\mu}^{s, \tau}\ k_{n^{\prime},\mu^{\prime}}^{s^{\prime}, \tau^{\prime}}}{\varepsilon_{n,d_{4}}  }     \Big)\delta_{n,n^{\prime}-1} 
-  \Big(    \frac{ \sqrt{n} 
}{  \ \varepsilon_{n,d_{2}}   }  +\frac{\sqrt{n^{\prime}+1}k_{n,\mu}^{s, \tau}\ k_{n^{\prime},\mu^{\prime}}^{s^{\prime}, \tau^{\prime}}}{\varepsilon_{n,d_{4}}^{\prime}  }     \Big)\delta_{n,n^{\prime}+1} \Big],\label{e22}
\end{multline}
where $\mu=\left\{  \mu_{1},\mu_{2}\right\}  $. 
  Using Eqs. (\ref{e19}), (\ref{e21}), and (\ref{e22}) we obtain
\begin{align}
\begin{split}
&\sigma_{yx}=\frac{e^{2}}{2h} 
\sum_{s,\tau,
\mu,\mu^{\prime}}\sum_{n}\Big[
\eta_{n,\mu,\mu^{\prime}}^{s, \tau}\ \frac{
f_{n,\mu}^{s, \tau}-f_{n+1,\mu^{\prime}}^{s, \tau}
}{\bigl(  \varepsilon_{n,\mu}^{s, \tau}-\varepsilon_{n+1,\mu^{\prime}}^{s, \tau
}\bigr)  ^{2}}  -\varsigma_{n,\mu,\mu^{\prime}}^{s, \tau
}\ \frac{
f_{n,\mu}^{s, \tau}-f_{n-1,\mu^{\prime}}^{s, \tau}
 }{\bigl(  \varepsilon_{n,\mu}^{s, \tau}-\varepsilon_{n-1,\mu^{\prime}}
^{s, \tau}\bigr)  ^{2}}  \Big], \label{e23}
\end{split}
\end{align}
with
\begin{align}
\begin{split}
&\eta_{n,\mu,\mu^{\prime}}^{s, \tau}=(
n+1)    \big(  \varrho_{n,\mu}^{s, \tau}\varrho_{n+1,\mu^{\prime}}^{s, \tau}\big)^{2} \,  \Big[  \frac{
    k_{n,\mu}^{s, \tau}\ k_{n+1,\mu^{\prime}}^{s, \tau}}      
  {  \ \varepsilon_{n,d_{4}}     }          +         \frac{1}{\varepsilon_{n+1,d_{2}}  } \Big]  ^{2},\label{e24}
\end{split}
\end{align}
\begin{align}
\begin{split}
&\varsigma_{n,\mu,\mu^{\prime}}^{s, \tau}=
n    \big(  \varrho_{n,\mu}^{s, \tau}\varrho_{n-1,\mu^{\prime}}^{s, \tau}\big)^{2} \,  \Big[  \frac{
    k_{n,\mu}^{s, \tau}\ k_{n-1,\mu^{\prime}}^{s, \tau}}      
  {  \ \varepsilon_{n-1,d_{4}}     }          +         \frac{1}{\varepsilon_{n,d_{2}}  } \Big]  ^{2}.\label{e25}
\end{split}
\end{align}
\end{widetext}
\begin{figure}[t]
\includegraphics[width=.49\textwidth]{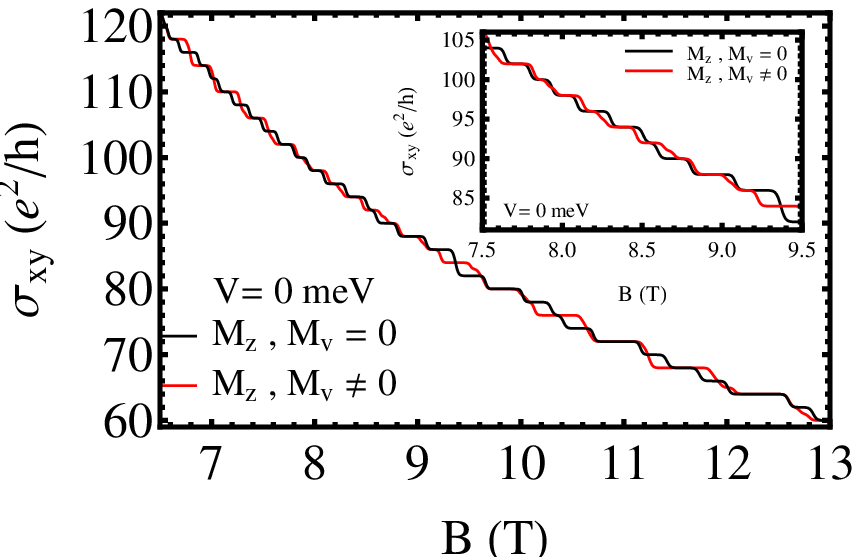}
\includegraphics[width=.49\textwidth]{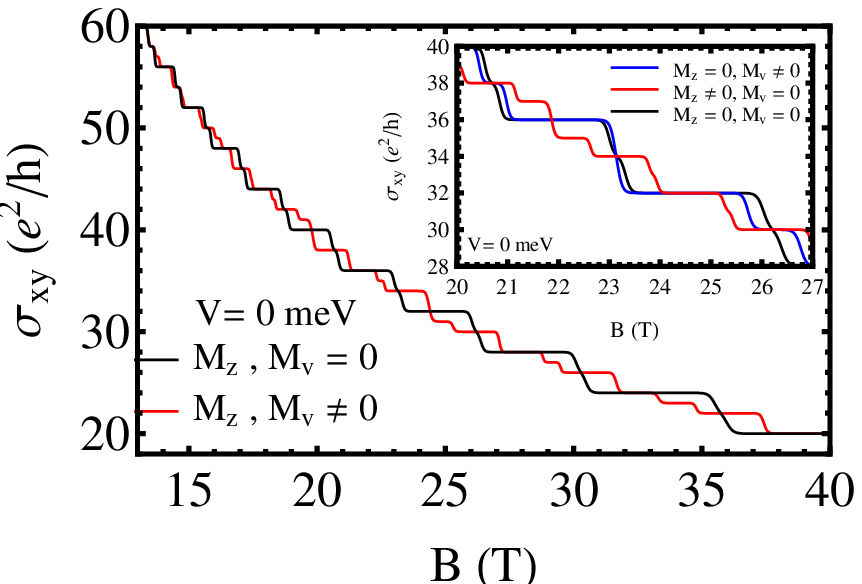}
\vspace{-0.5cm}
\caption{Hall conductivity as a function of the magnetic field $B$ for $T= 1$ K and $V=0$ meV. The two panels differ only in the  range of $B$. For further clarity, the range 7.5 T-9.5 T is shown in the inset to the left panel and the range 20 T-27 T in that to the right panel. }
\end{figure}
%
 The second term in Eq. (\ref{e23}) is valid only for $n\geq2$ while the first term is valid for $n\geq1$. This is so because the sum over $n$ is split in two parts, one for $n\geq1$ and one for $n=0$. Replacing $n-1$ with $n$ in the second term and combining it with the first term, 
 the sum over $n$ starts at $n=1$ for both the terms. 
 The $n =0$ 
 contribution to the Hall conductivity  Eq. (\ref{e23}) is evaluated separately using the eigenstates (\ref{e12}). The result is given by Eq. (\ref{A}) in  Appendix A. Furthermore, for the $n\geq1$ LLs  occupied, at $T= 0$, the $n=0$  LL contribution to the Hall conductivity  vanishes because all Fermi factors are equal to $1$. In the limit $V=\Delta=\lambda=0$,  Eq. (\ref{e23}) reduces to similar ones 
 for bilayer graphene \cite{17, 18}.

Figure 10 shows the Hall conductivity as a function of the field $B$ for $V=0$ meV. We found that the height of the steps is not constant: there are two different 
heights: $2\,e^{2}/h$ and $4\,e^{2}/h$
 see Fig. 10, black curve, in the absence of the spin and valley Zeeman terms. However, additional new heights 
 $2\,e^{2}/h$, $3\,e^{2}/h$ and $4\,e^{2}/h$ emerge in the sequence ladder in their presence 
as the red curve shows. These differences 
result from vanishing spin splittings as discussed in detail below  Eq. (\ref{e16}). Further, 
the plateaux in bilayer MoS$_{2}$ have different origin than those in bilayer graphene:  the former  are due to the  strong SOC whereas the later  result from strong interlayer coupling  \cite{17,18}. A noteworthy feature of  bilayer MoS$_{2}$ is that the 
influence  of SOC and interlayer  coupling is enhanced with increasing LL index 
and leads to 
new Hall plateaux as is evident from both panels of Fig. 10. In contrast to  monolayer MoS$_{2}$ \cite{9i}, the plateaux in  bilayer MoS$_2$ occur at higher magnetic fields.
\begin{figure}[t]
\includegraphics[width=.49\textwidth]{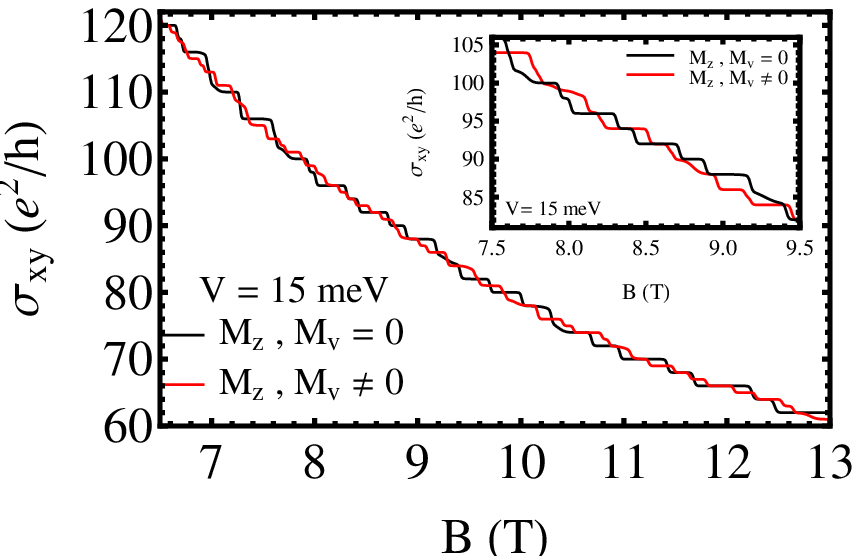}
\includegraphics[width=.49\textwidth]{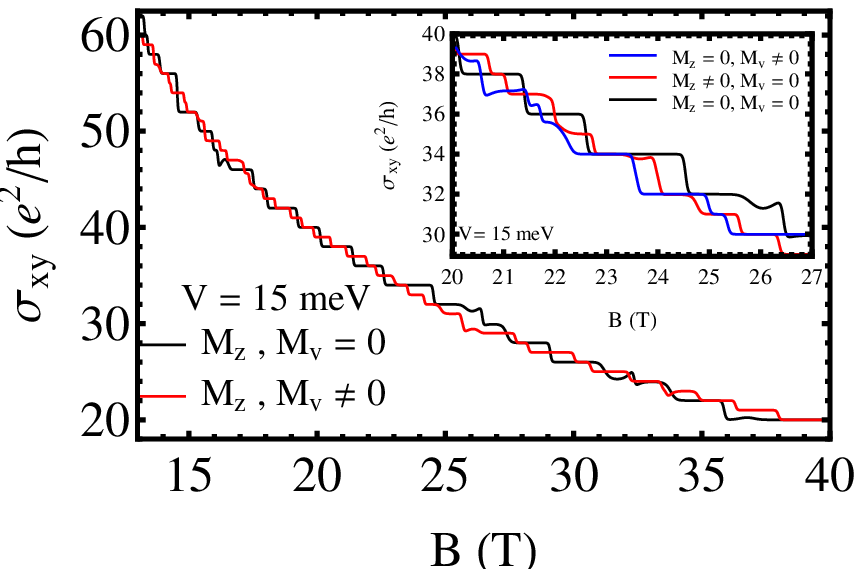}
\vspace{-0.5cm}
\caption{Hall conductivity as a function of the magnetic field for $T= 1$ K and $V=15$ meV. The two panels differ only in the  range of $B$ ($x$ axis). For further clarity, the range 7.5 T - 9.5 T is shown in the inset to the left panel and the range 20 T - 27 T in that to the right one.}
\end{figure}

We plot the Hall conductivity versus the  field $B$ in Fig. 11 for  electric field energy $V=15$ meV. For $M_{z}=M_{v}=0$ (black curve of Fig. 11), the plateaux appear at $0, 2, 4,...... (e^{2}/h)$. It is noted that new plateaux like four step size multiples of $e^{2}/h$ as seen in left and right panels of Fig. 11 (black curve) emerge at higher LLs due to level crossings  caused by the layer splitting. It is important to note that layer splitting is modifed by a finite  field $E_z$. On the other hand, additional plateaux emerge 
in the presence of spin and valley Zeeman fields, such as like $0,1, 2,...... (e^{2}/h)$. 
Interestingly, by contrasting 
Figs. $10$ and $11$  we see that the Hall plateau sequence strongly depends on the   field $E_z$. Furthermore, when $E_z$ is absent the plateaux occur at $0, 4, 8, 12,..... (e^{2}/h)$ as depicted in Fig. $10$ (black curve), whereas  for  a finite $E_z$, e. g., such that  $V=15$ meV, a new plateau sequence  emerges with a mixture of double and quadruple steps of integral multiples of $e^{2}/h$, such as $0, 2, 4, 6,..... (e^{2}/h)$ as shown in Fig. 11 (black curve). The latter is a result of layer splitting that is modified by the  field $E_z$. The emergence of new steps in the Hall conductivity 
is directly connected to the small jumps in the Fermi level as shown by the purple curves in Figs. 6 and 7. Importantly, at higher $B$ we find new plateaux in the Hall conductivity due to the spin and valley Zeeman fields in the absence and presence of the  field $E_z$ as the insets of Figs. 10 and 11 show.

\subsection{Collisional conductivity}

We assume that the electrons are elastically scattered by randomly distributed charged impurities. 
This type of scattering is dominant at low temperatures. If there is no spin degeneracy, the collisional conductivity is given by  \cite{24}
%
\begin{equation}
\sigma_{xx}=\frac{\beta e^{2}}{L_{x}L_{y}}\sum 
_{\zeta,\zeta^{^{\prime}}} f(  E_{\zeta})  \bigl[  1-f(
E_{\zeta^{^{\prime}}})  \bigr]   W_{\zeta\zeta^{^{\prime}}}
(  x_{\zeta}-x_{\zeta^{\prime}})  ^{2}.\label{e26}
\end{equation}
%
\noindent Here $f\left(  E_{\zeta}\right)  $ is the Fermi-Dirac 
function, 
$\beta=1/k_{B}T$,  $k_{B}$ is the Boltzmann constant, and $E_{F}$ the chemical potential. $W_{\zeta\zeta^{\prime}}$ is the transition rate between the one-electron states $\left\vert \zeta\right\rangle $ and $\left\vert\zeta^{\prime}\right\rangle $ and $e$  the  electron's charge. Conduction occurs by hopping between 
spatially separated states centered at $x_{\zeta}$ and $x_{\zeta^{\prime}}$,  $x_{\zeta}=\left\langle\zeta\right\vert x\left\vert \zeta\right\rangle $. 
\begin{figure}[t]
\centering
\includegraphics[width=.45\textwidth]{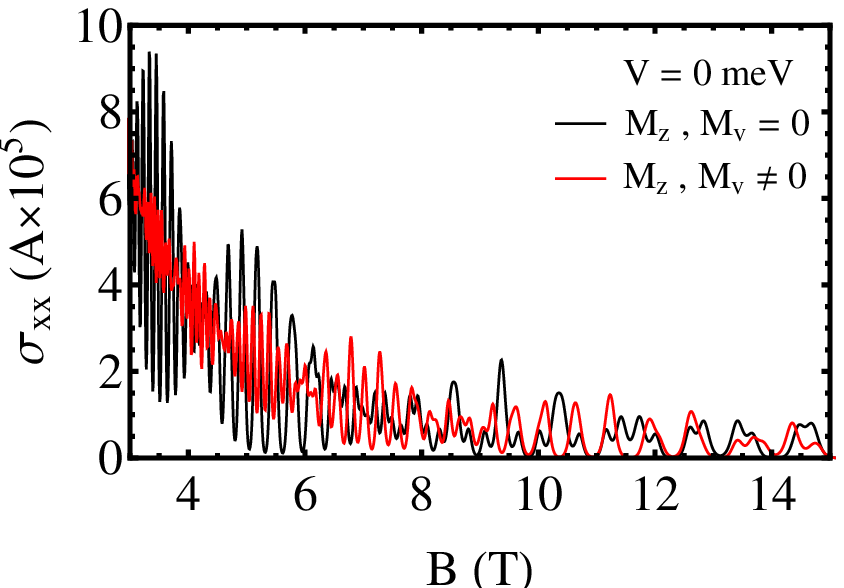}
\includegraphics[width=.45\textwidth]{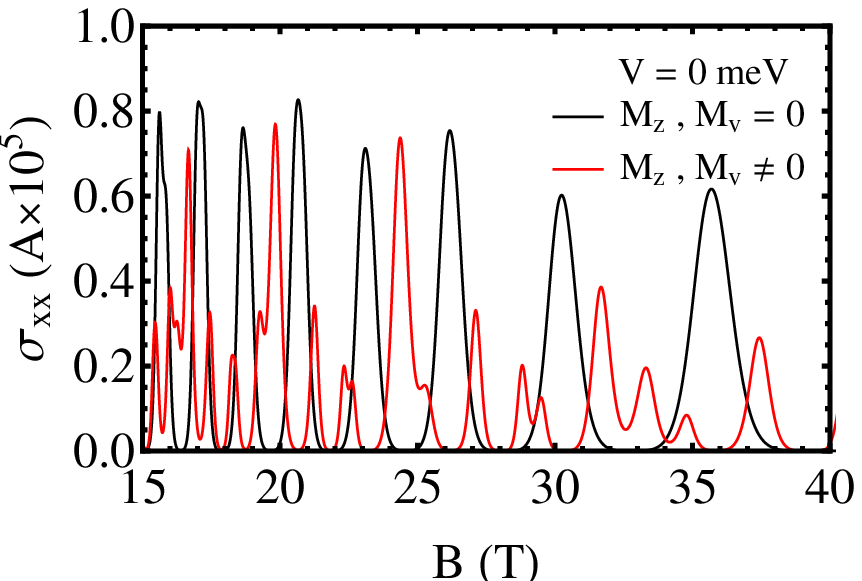}
\includegraphics[width=.45\textwidth]{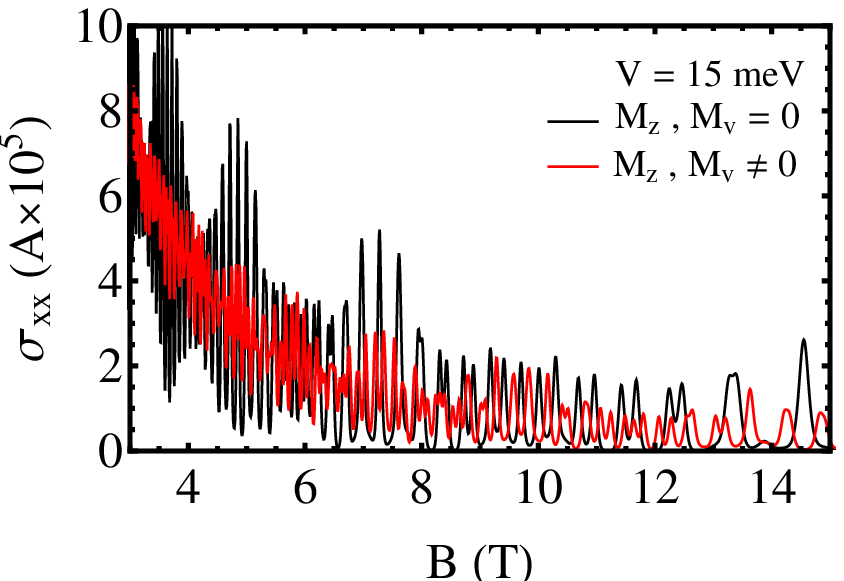}
\includegraphics[width=.45\textwidth]{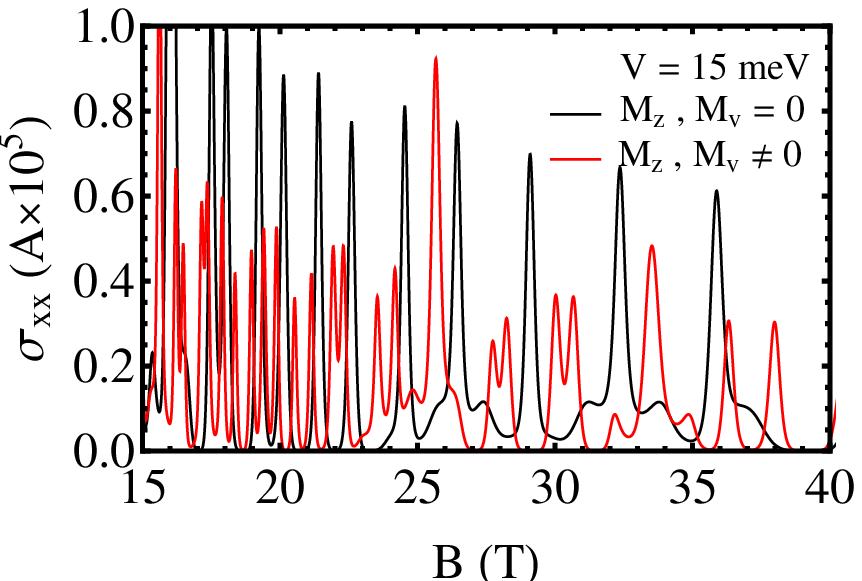}
\vspace{-0.5 cm}
\caption{Longitudinal conductivity versus magnetic field $B$ at $T= 1$ K. The upper (lower) panels are for $V=0$ meV ($V=15$ meV).
The left and right panels differ only in the  range of $B$.}
\end{figure}
The  rate $W_{\zeta\zeta^{\prime}}$ in Eq. (\ref{e26}) is given by
\begin{equation}
W_{\zeta\zeta^{\prime}}= \frac{2\pi N_{I}}{L_{x}L_{y}\hslash}\sum_{q} \left\vert U_{q}\right\vert
^{2}  \left\vert G_{\zeta\zeta^{\prime}}(r)\right\vert ^{2} \delta(
E_{\zeta}-E_{\zeta^{^{\prime}}})  \delta_{k_{y}^{\prime},k_{y}+q},\label{e27}
\end{equation}
with $q^2=q_{x}^{2}+q_{y}^{2}$ 
and $N_{I}$ the impurity density. 
For an  impurity at the origin the screened potential is given by $U({\bf r}
)=e^{2}e^{-k_{s}r} 
/\varepsilon r$ 
and its  Fourier transform   $U_{q}=U_{0}/ [q^{2}+k_{s}^{2}]^{1/2}$ with $U_{0}=2\pi e^{2}/\varepsilon$ and $k_{s}$ the screening wave vector. Further, if the impurity potential is short ranged, of the Dirac $\delta$-function type, one may use the approximation $k_{s}>>q$ and obtain $U_{q}\approx U_{0}/k_{s}$. $G_{\zeta\zeta^{\prime}}(r)=\left\langle \zeta^{\prime
}\right\vert e^{i\mathbf{q\cdot r}}\left\vert \zeta\right\rangle \ $ are the form factors and $\left\vert \zeta\right\rangle =\left\vert n,\mu,s,k_{y} \right\rangle $. Since the scattering by impurities is elastic and the spectrum is independent of $k_y$, we have $n=n^{\prime}$ and no LL mixing. 
Further, 
$\left(  x_{\zeta}-x_{\zeta^{\prime}}\right)^{2}=l_{B}^{4}q_{y}^{2}$. 
We notice that  the eigenfunction oscillates around the centre of the orbit $x_{0}=l_{B}^2k_{y}$  and make the changes $\sum_{k_{y}}\rightarrow (L_y/2\pi)\int_{-k_0}^{k_0}dk_{y}, k_{0}=L_{x}/2l_{B}^{2}$ 
and $\sum_{q}\rightarrow (L_{x}L_{y}/4\pi^{2}l_{B}^{2})\int_{0}^{2\pi}d\phi\int_{0}^{\infty}du$. The form factors $\left\vert \ G_{\zeta\zeta^{\prime}}(u)\right\vert ^{2}$ can be evaluated from the matrix element. For  $n'=n, s=s', \mu=\mu'$ we obtain 
\begin{equation}
\hspace*{-0.02cm}\left\vert G_{nn}(u)\right\vert ^{2}=e^{-u }\Big[
\big[  1+(k_{n,\mu}^{s, \tau}\bigr)^{2}  \big] L_{n}(u)  
+\frac{n}{  \varepsilon_{n,d_{2}}^{2} 
}\ L_{n-1}(u) 
+\frac{\left(  n+1\right)(k_{n,\mu}^{s,\tau}\bigr)^{2}  }{\varepsilon_{n,d_{4}}^{2}}\ L_{n+1}(u)
\Big]^{2},\label{e28}
\end{equation} 
%
 with  $u=l_{B}^{2}q^{2}/2$ and  $L_{n}(u)$ the associated Laguerre polynomials. 
 Inserting all form factors in Eq. (\ref{e26}) and evaluating the integral  over $u$ in cylindrical coordinates 
gives
\begin{eqnarray}
\notag
 \sigma_{xx} =A\sum_{n,\mu,s,\tau}
(  \varrho_{n,\mu}^{s,\tau}
)  ^{4} &\Big[ &(2n+1) 
 \big[ 1+\bigl(k_{n,\mu}^{s,\tau}\bigr)^{2}  \big]^{2} 
 +\frac{( 2n-1)n^{2}}{   \varepsilon_{n,d_{2}}^{4} } +\frac{(2n+3) ( n+1) ^{2} \bigl(k_{n,\mu}^{s,\tau}\bigr)^{4}}{\varepsilon_{n,d_{4}}^{4}} \Big] \\*
&\times& f(E_{n,\mu}^{s,\tau})[1-f(E_{n,\mu}^{s,\tau})],\label{e29}
\end{eqnarray}
where 
$A= (e^{2}/h)(\beta N_{I}\left\vert U_{0}\right\vert ^{2}/\pi l_{B}^{2}\Gamma k_{s}^{2}$ and $\Gamma$ is the level width. Note that Eq. (\ref{e29}) reduces to that 
for bilayer graphene \cite{17} in the limit $V=\Delta=\lambda=0$. The collisional conductivity for 
 $n=-1,0$  is given in Appendix B. 
\begin{figure}[t]
\centering
\includegraphics[width=.57\textwidth]{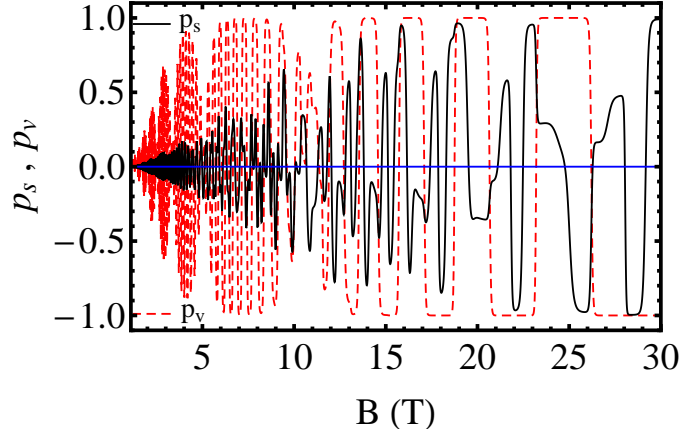}
\vspace{-0.5 cm}
\caption{Spin $P_{s}$ and valley $P_{v}$ polarizations versus magnetic field $B$ at $T= 1$ K. The parameters are the same as in Fig. 11 for $M_{z}\neq M_{v}\neq0$.}
\end{figure}

 The longitudinal conductivity $\sigma_{xx}$, given by Eq. (\ref{e29}), is shown  in Fig. 12 as a function of the  field $B$ for $E_z=0$ (upper panels) and $E_z$ finite (lower panels). In contrast to bilayer graphene, Fig. 12 shows a beating pattern of the SdH oscillations for $B$ fields up to $9$ T when  $E_z$ is absent ($V=0$) and for $B$ fields up to $7$ T when a finite $E_z$ is present ($V=15$ meV). For  high $B$ fields the beating pattern is absent and the longitudinal conductivity peaks are split. The beating pattern is controlled by the  fields $E_z$ and $B$. A typical beating pattern occurs when the LL level broadening is of the same order as the LL separation. The SOC  becomes weak at larger $B$ fields. 
Interestingly, in contrast to monolayer MoS$_{2}$ \cite{9i}, the position of the nodes depends on both the field $E_z$ and spin and valley Zeeman terms. The results of the collisional conductivity are consistent with the Fermi energy and DOS as seen in Figs. 8-9. Analytically, the beating of the SdH oscillations can be understood  by making the approximation $\beta f(E_{n,\mu}^{s,\tau})[1-f(E_{n,\mu}^{s,\tau})]\approx \delta(E_{F}-E_{n,\mu}^{s,\tau})$ at very low temperatures in Eq. (\ref{e29}), broadening the delta function, and carrying out the procedure  followed in Ref. \onlinecite{9i}.

 The spin $P_{s}$ and valley $P_{v}$ polarization, which are extracted from Eq. (\ref{e29}), are
\begin{equation}
P_{s}=\dfrac{(\sigma_{xx}^{K,\uparrow}+\sigma_{xx}^{K^{\prime},\downarrow})-(\sigma_{xx}^{K,\downarrow}+\sigma_{xx}^{K^{\prime},\uparrow})} {(\sigma_{xx}^{K,\uparrow}+\sigma_{xx}^{K^{\prime},\downarrow})+(\sigma_{xx}^{K,\downarrow}+\sigma_{xx}^{K^{\prime},\uparrow})},\label{e30}
\end{equation}
and
\begin{equation}
P_{v}=\dfrac{(\sigma_{xx}^{K,\uparrow}+\sigma_{xx}^{K,\downarrow})-(\sigma_{xx}^{K^{\prime},\uparrow}+\sigma_{xx}^{K^{\prime},\downarrow})} {(\sigma_{xx}^{K,\uparrow}+\sigma_{xx}^{K,\downarrow})+(\sigma_{xx}^{K^{\prime},\uparrow}+\sigma_{xx}^{K^{\prime},\downarrow})}.\label{e31}
\end{equation}
\begin{figure}[t]
\centering
\includegraphics[width=.45\textwidth]{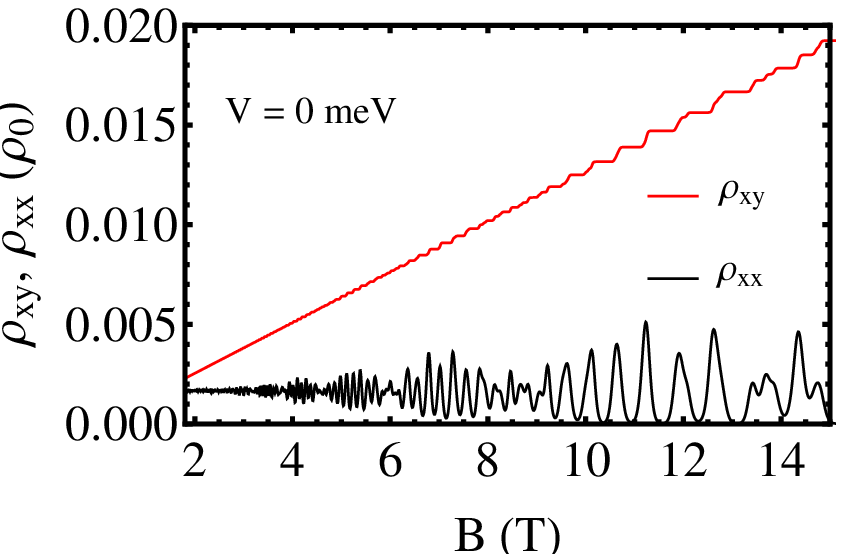}
\includegraphics[width=.45\textwidth]{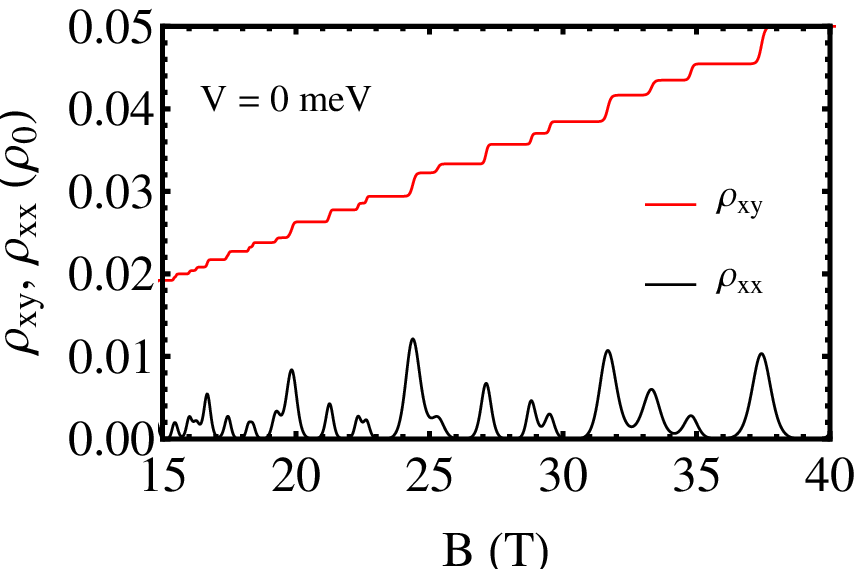}
\includegraphics[width=.45\textwidth]{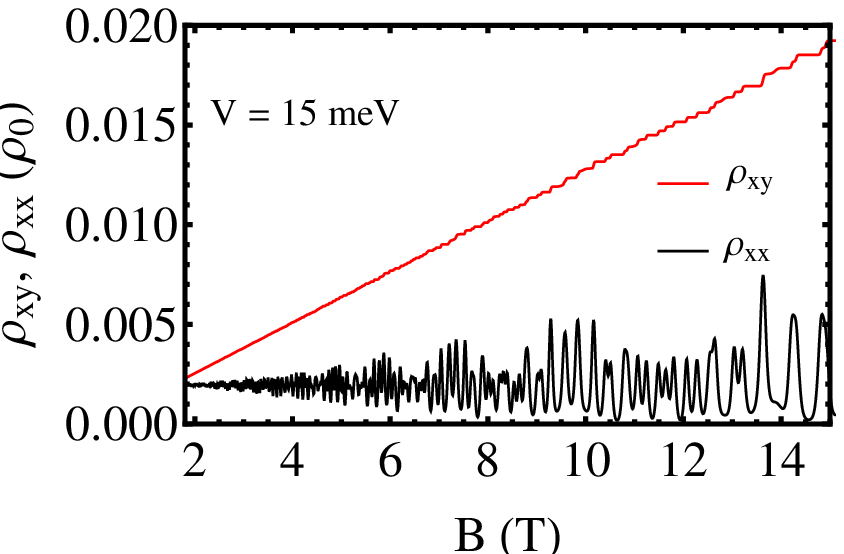}
\includegraphics[width=.45\textwidth]{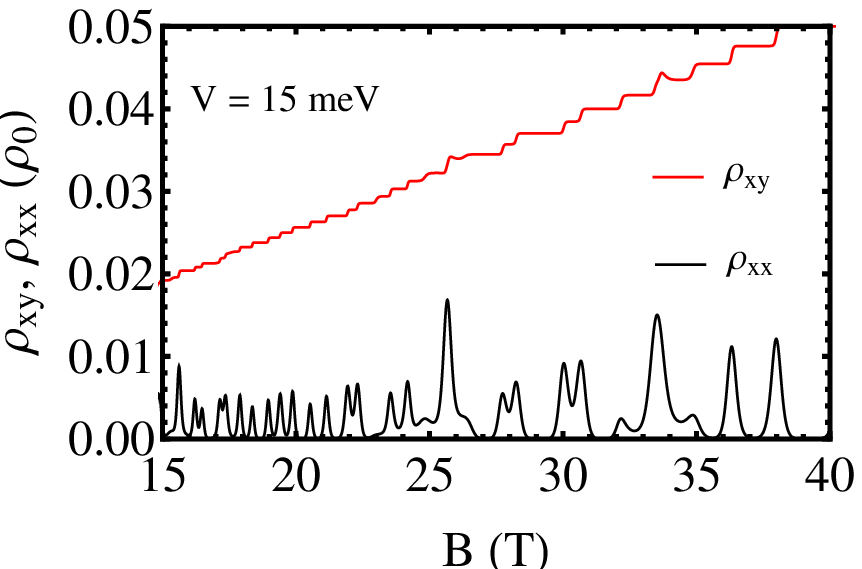}
\vspace{-0.5 cm}
\caption{Longitudinal (black) and Hall (red) resistivities versus magnetic field $B$ at $T= 1$ K and finite spin and valley Zeeman fields. The upper panels are for $V=0$ meV and the lower ones for $V=15$ meV. The left and right panels differ only in the  range of $B$ and $\rho_{0}=A^{-1}\times 10^{-35}$.}
\end{figure}
We plot the spin $P_{s}$ (black solid curve) and $P_{v}$ (red dotted curve) polarization versus magnetic field at $T=1$ K, $V=0$ meV and finite Zeeman fields in Fig. 13. As expected and can be seen, here too we have a beating pattern at low magnetic fields and 
well-resolved separation between both $P_{s}$ and $P_{v}$ at higher magnetic fields. The fact is that strong magnetic fields give rise to larger splittings of the LLs. In contrast to monolayer MoS$_{2}$ \cite{9i}, we find $100\%$ valley polarization above $B>13$ T whereas we attain $90\%$ spin polarization above $B>20$ T. Notice also the square-wave character of $P_{v}$ above $B>13$ T. However, for $M_{z}=M_{v}=0$, there is no $P_{s}$ and $P_{v}$ as shown by the blue curve.

 Finally, we evaluate the magnetoresistivity $\rho_{\mu\nu}$ using the conductivity tensor via the well-known relations  $\rho_{xx}=\sigma_{xx}/S$ and $\rho_{xy}=\sigma_{xy}/S$ with $S=\sigma_{xx}\sigma_{yy}-\sigma_{xy}\sigma_{yx}\approx n_{e}^{2}e^{2}/B^{2}$ where $n_{e}$ is the electron concentration. The Hall and longitudinal resistivities are shown in Fig. 14 versus magnetic field $B$ for $T=1$ K with 
field energy $V=0$ meV (upper   panels)  $V=15$ meV ( lower panels). We observe extra plateaux in the Hall resistivity due to the SOC, layer splitting, and spin and valley Zeeman terms. The steps between the plateaux coincide with sharp peaks in the longitudinal resistivity. For $V=0$ meV and strong $B$ fields, larger than $13$ T, we find a significant splitting of the Hall plateaux and the corresponding peaks in the longitudinal resistivity due to spin and valley Zeeman fields. On the other hand, for $V=15$ meV and $B$ fields larger than $8$ T, we find a well-resolved splitting of the Hall plateaux and the corresponding peaks of the longitudinal resistivity due to spin and valley Zeeman terms and inter-layer splitting. In contrast, for $B$ fields less than $13$ T ($V=0$ meV) and $8$ T ($V=15$ meV),  the longitudinal resistivities show a beating pattern. Importantly, this  pattern is similar to that in a conventional 2DEG in the presence of the Rashba SOC \cite{i3}. Also, we note that well-resolved plateaux occur at relatively higher $B$ than in monolayer MoS$_{2}$ \cite{9i}. 
We expect that these results will be verified by experiments. 

\vspace{-.5cm}
\section{conclusions}

 We  studied  quantum magnetotransport in bilayer MoS$_{2}$ in the presence of  perpendicular electric ($E_z$) and magnetic ($B$) fields. At $B=0$ we showed that  there is no spin splitting 
for zero field $E_{z}$ in both the conduction and valence bands whereas there is one for 
finite field $E_{z}$. Further, for $E_{z}\neq 0$ we demonstrated that the conduction band is still spin degenerate while the spin degeneracy in the valence band is fully lifted (see Fig.~1). We showed though that the layer splitting and band gap can be controlled by the 
 field $E_{z}$. 
  The spin degeneracy of the levels, for  $E_{z}=0$, in the conduction band,  is lifted 
 for $B\neq0$ and is also enhanced  linearly with 
 $B$ (see text after Eq. (15)). Furthermore, a finite field $E_{z}$ leads to a significant enhancement of the spin splitting energy in the adjacent LLs of the conduction band. For $V=0$ meV ($V=15$ meV) and $B\leq13$ T ($B\leq8$ T), the Fermi energy $E_{F}$ and DOS show a 
 beating pattern which 
 is replaced by a split of the SdH oscillations above $B>13$ T ($B>8$ T). Moreover, we 
showed that the combined action of spin and valley Zeeman fields and inter-layer splitting allow for intra-LL transitions and lead to new quantum Hall plateaux. 
The 
field $E_z$ 
  modifies the layer splitting. As a result, 
steps  of various heights, in multiples of $e^2/h$ (Fig. 11), occur in the Hall conductivity.
 Furthermore, for $V=0$ meV ($V=15$ meV) and $B>9$ T ($B>7$ T) the number of peaks in the longitudinal conductivity is doubled whereas for fields $B<9$ T ($B<7$ T) a beating pattern occurs similar to monolayer MoS$_{2}$ \cite{9i} and the conventional 2DEG \cite{i3}. 

Beating patterns, at low $B$ fields, and splittings, at strong $B$ fields, also occur in the spin and valley polarizations. It is  worth emphasizing that a $100\%$, 
 square-wave-shaped  valley polarization is obtained for $B>13$ T and $90\%$ spin polarization for $B>20$ T. The deep minima in the SdH oscillations are accompanied by Fermi level jumps and the peaks coincide with the usual singularities of the DOS. A beating pattern and splitting of the SdH oscillations occur also in the resisitivity that can be controlled by the magnetic field $B$ which 
enhances the spin splitting in the conduction band. 
The spin and valley Zeeman fields lead to a giant splitting for strong $B$ fields and to a lifting of the fourfold spin and valley degeneracies. The position of the plateaux as well as the peaks and beating pattern are sensitive to  the field $E_z$ and to the spin and valley Zeeman fields. 
 The latter increase the number of beating nodes in the longitudinal conductivity, $E_{F}$, and DOS.  The results, which  we hope  will be tested by experiments, indicate that bilayer MoS$_{2}$ is a promising alternative to bilayer graphene in the quest for gapped Dirac materials. We expect further applications of bilayer MoS$_{2}$ 
in the field of valleytronics and spintronics. 

\vspace{0.3cm}
{\bf Acknowledgments}: 
M. Z. and K. S. acknowledge the support of Higher Education Commission of Pakistan through project No. $20-1484$/R$\&$D$/09$ . K. S. also  acknowledges the support of the Abdus Salam International Center for Theoretical Physics (ICTP) in Trieste, Italy, through the Associate Scheme where part of this work was completed. This work was supported by the the University of Hafr Al Batin (MT). The work of P. V. was supported by the Canadian NSERC Grant No. OGP0121756.

${*}$ (m.tahir06@alumni.imperial.ac.uk and tahir@uohb.edu.sa)

\appendix
\section{Zero-level Hall conductivity}

Using Eqs. (\ref{e12}) the off-diagonal velocity matrix elements for  $n=0$ are 
\begin{multline}
\left\langle 0,\mu,s,\tau\right\vert v_{x}\left\vert n^{\prime},\mu^{\prime
},s^{\prime},\tau^{\prime}\right\rangle   =\tau v_{F }\varrho_{0,\mu}^{s,\tau
}\varrho_{n^{\prime},\mu^{\prime}}^{s^{\prime}, \tau^{\prime}} \delta_{s,s^{\prime}}\\
  \times\Big[
\Big\{\frac{\sqrt{n^{\prime}}}{\varepsilon_{n,d_{2} }^{\prime} }   +\frac{k_{0,\mu}^{s,\tau} k_{n^{\prime},\mu^{\prime}}^{s^{\prime},\tau^{\prime}} } {\varepsilon_{0,d_{4}} }   \Big\} \ \delta_{0,n^{\prime}-1}   
+  \frac{\sqrt{n^{\prime}+1}k_{0,\mu}^{s,\tau}k_{n^{\prime},\mu^{\prime}}^{s^{\prime},\tau^{\prime}} }{\varepsilon_{n,d_{4}}^\prime
  } 
 \ \delta_{0,n^{\prime}+1} \Big],
\end{multline}
\begin{multline}
\left\langle n^{\prime},\mu^{\prime},s^{\prime},\tau^{\prime}\right\vert
v_{y}\left\vert 0,\mu,s,\tau\right\rangle    =\tau i v_{F }\varrho_{0,\mu}^{s,\tau
}\varrho_{n^{\prime},\mu^{\prime}}^{s^{\prime}, \tau^{\prime}} \delta_{s,s^{\prime}}\\
  \times\Big[
\Big\{\frac{\sqrt{n^{\prime}}}{\varepsilon_{n,d_{2} }^{\prime} }   +\frac{k_{0,\mu}^{s,\tau} k_{n^{\prime},\mu^{\prime}}^{s^{\prime},\tau^{\prime}} } {\varepsilon_{0,d_{4}} }   \Big\} \ \delta_{0,n^{\prime}-1}   
-  \frac{\sqrt{n^{\prime}+1}k_{0,\mu}^{s,\tau}k_{n^{\prime},\mu^{\prime}}^{s^{\prime},\tau^{\prime}} }{\varepsilon_{n,d_{4}}^\prime
  } 
 \ \delta_{0,n^{\prime}+1} \Big].
\end{multline}
%
%
For $n=-1$  
we find
\begin{equation}
\left\langle -1\right\vert v_{x}\left\vert n^{\prime},\mu^{\prime
},s^{\prime},\tau^{\prime}\right\rangle =\tau v_{F}\varrho_{n^{\prime},\mu^{\prime}%
}^{s^{\prime}, \tau^{\prime}} k_{n^{\prime},\mu^{\prime}}^{s^{\prime},\tau^{\prime}}\delta_{s,s^{\prime}}\, \delta_{0,n^{\prime}},
\end{equation}
\begin{equation}
\left\langle n^{\prime},\mu^{\prime},s^{\prime},\tau^{\prime}\right\vert
v_{y}\left\vert -1 \right\rangle =\tau iv_{F}\varrho_{n^{\prime},\mu^{\prime}%
}^{s^{\prime}, \tau^{\prime}} k_{n^{\prime},\mu^{\prime}}^{s^{\prime},\tau^{\prime}}\delta_{s,s^{\prime}}\,\delta_{0,n^{\prime}},
\end{equation}
Using these expressions the Hall conductivity takes the form
\begin{equation}
\sigma_{yx}    =\frac{e^{2}}{h}
\sum_{s,\tau}\sum_{\mu,\mu^{\prime}
}  \left[ \eta_{0,1,\mu,\mu^{\prime}}^{s,\tau}
\frac{
f_{0,\mu}^{s, \tau}-f_{1,\mu^{\prime}}^{s,\tau
}
 }{\bigl(  \varepsilon_{0,\mu}^{s, \tau}-\varepsilon_{1,\mu^{\prime}}^{s, \tau
}\bigr)  ^{2}}  
+\bigl(  \varrho_{0,\mu^{\prime}}
^{s, \tau} k_{0,\mu^{\prime}}^{s,\tau} \bigr)  ^{2}\frac{
f_{-1}^{s, \tau
}-f_{0,\mu^{\prime}}^{s, \tau}
 }{\bigl(
\varepsilon_{-1}^{s,\tau}-\varepsilon_{0,\mu^{\prime}}^{s, \tau}\bigr)  ^{2}
}\right],\label{A}
\end{equation}
where
\begin{equation}
\eta_{0,1,\mu,\mu^{\prime}}^{s, \tau}=  \bigl(  \varrho_{0,\mu}^{s, \tau}\varrho_{1,\mu^{\prime}}^{s, \tau
}\bigr)  ^{2}  \Big[  \frac{1}{\varepsilon_{1,d_{2} }^{\prime} }   +\frac{k_{0,\mu}^{s,\tau} k_{1,\mu^{\prime}}^{s,\tau} } {\varepsilon_{0,d_{4}} }   \Big]^{2},
\end{equation}
%
%
\section{Zero-level collisional conductivity}

The form factors for  $n=0$  and $n=-1$, with  $n'=n, s=s'$, and $\mu=\mu'$, are given by 
\begin{equation}
\hspace*{-0.02cm}\left\vert G_{00}(u)\right\vert ^{2}=e^{-u }\Big[
\big[  1+(k_{0,\mu}^{s, \tau}\bigr)^{2}  \big] L_{0}(u)  
+\frac{(k_{0,\mu}^{s,\tau}\bigr)^{2}  }{\varepsilon_{0,d_{4}}^{2}}\ L_{1}(u)
\Big]^{2},
\end{equation} 
and
\begin{equation}
\hspace*{-0.02cm}\left\vert G_{-1-1}(u)\right\vert ^{2}=e^{-u }L_{0}^{2}(u)
\end{equation}

The collisional conductivity is 

\begin{eqnarray}
\notag
 \sigma_{xx} =A\sum_{\mu,s,\tau}
\Big[
(  \varrho_{0,\mu}^{s,\tau}
)  ^{4} &\Big[ &
 \big[ 1+\bigl(k_{0,\mu}^{s,\tau}\bigr)^{2}  \big]^{2} 
 +\frac{3 \bigl(k_{0,\mu}^{s,\tau}\bigr)^{4}}{\varepsilon_{0,d_{4}}^{4}} \Big]  f(E_{0,\mu}^{s,\tau})[1-f(E_{0,\mu}^{s,\tau})]\\*
&+& f(E_{-1}^{s,\tau})[1-f(E_{-1}^{s,\tau})]\Big],
\end{eqnarray}


\begin{thebibliography}{99}

\bibitem{5i}  D. Xiao, G. Liu, W. Feng, X. Xu, and W. Yao, Phys. Rev. Lett. {\bf 108}, 196802 (2012).

\bibitem{1i} A. Ayari, E. Cobas, O. Ogundadegbe, and M. S. Fuhrer, J. Appl. Phys. {\bf 101}, 014507 (2007).

\bibitem{2i} K. Mak, C. Lee, J. Hone, J. Shan, and T. Heinz, Phys. Rev. Lett. {\bf 105}, 136805 (2010).

\bibitem{3i} B. Radisavljevic, A. Radenovic, J. Brivio, V. Giacometti, and A. Kis, Nat. Nanotech. {\bf 6}, 147 (2011).

\bibitem{4i} A. Splendiani, L. Sun, Y. Zhang, T. Li, J. Kim, C.-Y. Chim, G. Galli, and F. Wang, Nano Lett. {\bf 10}, 1271 (2010).


\bibitem{7ii} X. Zhou, Y. Liu, M. Zhou, H. H. Shao, and G. H. Zhou, Appl. Phys. Express {\bf 7} 021201 (2014).

\bibitem{8i} F. Rose, M. O. Goerbig, and F. Piechon, Phys. Rev. B {\bf 88}, 125438 (2013).

\bibitem{8ii} R.-L. Chu, X. Li, S. Wu, Q. Niu, W. Yao, X. Xu, and C. Zhang, Phys. Rev. B {\bf 90}, 045427 (2014).

\bibitem{8iii} Y.-H. Ho, Y.-H. Wang, and H.-Y. Chen, Phys. Rev. B {\bf 89}, 155316 (2014).

\bibitem{4} X. Li, F. Zhang, and Q. Niu, Phys. Rev. Lett. {\bf 110}, 066803 (2013).

\bibitem{5} X. Zhou, Y. Liu, M. Zhou, D. Tang, and G. Zhou, J. Phys.: Condens. Matter {\bf 26}, 485008 (2014).

\bibitem{9i} M. Tahir, P. Vasilopoulos, and F. M. Peeters, Phys. Rev. B {\bf 93}, 035406 (2016).

\bibitem{z8} A. Korm\'anyos, P. Rakyta, and G. Burkard, New. J. Phys. {\bf 17}, 103006 (2015).

\bibitem{6} Q. Liu, L. Li, Y. Li, Z. Gao, Z. Chen, and J. Lu, J. Phys. Chem. C {\bf 116}, 21556 (2012).

\bibitem{7} A. Ramasubramaniam, D. Naveh, and E. Towe, Phys. Rev. B {\bf 84}, 205325 (2011).

\bibitem{8} N. Zibouche, P. Philipsen, A. Kuc, and T. Heine, Phys. Rev. B {\bf 90}, 125440 (2014).

\bibitem{9} Z. Gong, G.-B. Liu, H. Yu, D. Xiao, X. Cui, X. Xu, and Wang Yao, Nat. Commu. {\bf 4}, 15 (2013).

\bibitem{10} S. Wu, J. S. Ross,	G. B. Liu, G. Aivazian, A. Jones, Z. Fei, W. Zhu, D. Xiao, W. Yao, D. Cobden, and X. Xu, Nat. Phys. {\bf 9}, 149 (2013).

\bibitem{11} J. Lee, K. F. Mak, and J. Shan, Nat. Nanotech. (2016).doi:10.1038/nnano.2015.337.

\bibitem{13} A. T. Neal, H. Liu, J. J. Gu, and P. D. Ye, ACS Nano {\bf 7}, 7077 (2013).

\bibitem{14} F. Guinea, New J. Phys. {\bf 12}, 083063 (2010).

\bibitem{15} F. Mireles and J. Schliemann, New J. Phys. {\bf 14}, 093026 (2012).

\bibitem{16} J. Milton Pereira, Jr., P. Vasilopoulos, and F. M. Peeters, Phys. Rev. B {\bf 76}, 115419 (2007).

\bibitem{17} M. Zarenia, P. Vasilopoulos, and F. M. Peeters, Phys. Rev. B {\bf 85}, 245426 (2012).

\bibitem{18} M. Nakamura, L. Hirasawa, and K. I. Imura, Phys. Rev. B {\bf 78}, 033403 (2008).

\bibitem{19} K. Lee, S. Kim, M. S. Points, T. E. Beechem, T. Ohta, and E. Tutuc, Nano Lett. {\bf 11}, 3624 (2011).
\bibitem{20} M. A. Hidalgo, and R. Cangas, arXiv:1602.02631.

\bibitem{21} K. S. Novoselov, E. McCann, S. V. Morozov, V. I. Falko, M. I. Katsnelson, U. Zeitler, D. Jiang, F. Schedin, and A. K. Geim, Nat. Phys. {\bf 2}, 177 - 180 (2006). 

\bibitem{22} C. R. Dean, A. F. Young, I. Meric,	C. Lee,	L. Wang, S. Sorgenfrei,	K. Watanabe, T. Taniguchi, P. Kim,	K. L. Shepard, and J. Hone, Nat. Nanotech. {\bf 5}, 722� ��726 (2010).

\bibitem{1b} E. McCann, Phys. Rev. B {\bf 74}, 161403 (2006).

\bibitem{2b} Y. Zhang, T.-T. Tang, C. Girit, Z. Hao, M. C. Martin, A. Zettl, M. F. Crommie, Y. R. Shen, and F. Wang, Nat. Phys. {\bf 459}, 820 (2009).

\bibitem{3b} T. Ohta, A. Bostwick, T. Seyller, K. Horn, and E. Rotenberg, Science {\bf 313}, 951 (2006).

\bibitem{4b} F. Xia, D. B. Farmer, Y. Lin, and P. Avouris, Nano Lett. {\bf 10}, 715 (2010).

\bibitem{23} X. Cui, G.-H. Lee, Y. D. Kim, G. Arefe, P. Y. Huang, C.-H. Lee, D. A. Chenet, X. Zhang, L. Wang, F. Ye, F. Pizzocchero, B. S. Jessen, K. Watanabe, T. Taniguchi, D. A. Muller, T. Low, P. Kim, and J. Hone, Nat. Nanotechnol. {\bf 10}, 534 (2015).

\bibitem{24} M. Charbonneau, K. M. Van Vliet, and P. Vasilopoulos, J. Math. Phys. {\bf 23}, 318 (1982).

\bibitem{25}  A. M. Jones, H. Yu, J. S. Ross, P. Klement, N. J. Ghimire, J. Yan,	D. G. Mandrus, W. Yao and X. Xu, Nat. Phys.  {\bf 10}, 130 (2014).

\bibitem{7i} S. Fang, R. K. Defo, S. N. Shirodkar, S. Lieu, G. A. Tritsaris, and E. Kaxiras, Phys. Rev. B  {\bf 92}, 205108 (2015).

\bibitem{z1} A. Korm\'anyos, V. Z\'olyomi, N. D. Drummond, and G. Burkard, Phys. Rev. X {\bf 4}, 011034 (2014).

\bibitem{z2} D. MacNeill, C. Heikes, K. F. Mak, Z. Anderson, A. Korm\'anyos, V. Z\'olyomi, J. Park, and D. C. Ralph, Phys. Rev. Lett. {\bf 114}, 037401 (2015).

\bibitem{z3} A. Srivastava, M. Sidler, A. V. Allain, D. S. Lembke, A. Kis, and A. Imamoglu, Nat. Phys. {\bf 11}, 141 (2015).

\bibitem{z4} G. Aivazian, Z. Gong, A. M. Jones, R.-L. Chu, J. Yan, D. G. Mandrus, C. Zhang, D. Cobden, W. Yao, and X. Xu, Nat. Phys. {\bf 11}, 148 (2015).

\bibitem{z5} Y. Li, J. Ludwig, T. Low, A. Chernikov, X. Cui, G. Arefe, Y. D. Kim, A. M. van der Zande, A. Rigosi, H. M. Hill, S. H. Kim, J. Hone, Z. Li, D. Smirnov, and T. F. Heinz, Phys. Rev. Lett. {\bf 113}, 266804 (2014).

\bibitem{z6} Y. C. Cheng, Q. Y. Zhang, and U. Schwingenschl\"ogl, Phys. Rev. B {\bf 89}, 155429 (2014).

\bibitem{26} M. Koshino and T. Ando, Phys. Rev. B {\bf 81}, 195431 (2010).

\bibitem{new1} T. Cheiwchanchamnangij and W. R. L. Lambrecht, Phys. Rev. B {\bf 85}, 205302 (2012).

\bibitem{new2} A. Kumar and P. K. Ahluwalia, Modelling Simul. Mater. Sci. Eng. {\bf 21}, 065015 (2013).

\bibitem{new3} W. Jin, P.-C. Yeh, N. Zaki, D. Zhang, J. T. Sadowski, A. Al-Mahboob, A. M. v. d. Zande, D. A. Chenet, J. I. Dadap, I. P. Herman, P. Sutter, J. Hone, and R. M. Osgood, Jr. Phys. Rev. Lett. {\bf 111}, 106801 (2013).

\bibitem{z7} P. Koskinen, I. Fampiou, and A. Ramasubramaniam, Phys. Rev. Lett. {\bf 112}, 186802 (2014).

\bibitem{9ii} P. M. Krstajic and P. Vasilopoulos, Phys. Rev. B  {\bf 83}, 075427 (2011); {\it ibid.}  {\bf 86} 115432 (2012)
\bibitem{i2} M. Tahir, A. Manchon, and U. Schwingenschl\"ogl, Phys. Rev. B  {\bf 90}, 125438 (2014).

\bibitem{i3} X. F. Wang and P. Vasilopoulos, Phys. Rev. B  {\bf 72}, 085344 (2005); {\it ibid.} {\bf 67}, 085313 (2003).


\end{thebibliography}
\end{document}